\documentclass[prd,preprint,showpacs,preprintnumbers,amsmath,amssymb,eqsecnum]{revtex4-1}

\usepackage{graphicx}
\usepackage{dcolumn}
\usepackage{bm}
\usepackage{color}
\usepackage{url}
\usepackage[dvipdfmx]{hyperref}

\newcommand{\bk}{\ensuremath{{\bf k}}}
\newcommand{\bx}{\ensuremath{{\bf x}}}
\newcommand{\by}{\ensuremath{{\bf y}}}

\begin{document}

\preprint{RUP-17-12}
\preprint{KEK-Cosmo-206}
\preprint{KEK-TH-1987}
\preprint{OCU-PHYS-466}
\preprint{AP-GR-139}

\title{Spins of primordial black holes formed in the 
matter-dominated phase of the Universe
}

\author{Tomohiro Harada}%
\email{harada@rikkyo.ac.jp}
\affiliation{Department of Physics, Rikkyo University, Toshima,
Tokyo 171-8501, Japan}
\author{$^{1}$Chul-Moon Yoo}
\affiliation{$^{1}$Gravity and Particle Cosmology Group,
Division of Particle and Astrophysical Science,
Graduate School of Science, Nagoya University, Nagoya 464-8602, Japan}
\author{$^{2,3,4}$Kazunori Kohri}
\affiliation{$^{2}$Institute of Particle and Nuclear Studies, KEK,
1-1 Oho, Tsukuba, Ibaraki 305-0801, Japan}
\affiliation{$^{3}$The Graduate University for Advanced Studies (SOKENDAI),
1-1 Oho, Tsukuba, Ibaraki 305-0801, Japan}
\affiliation{$^{4}$Rudolf Peierls Centre for Theoretical Physics, The University of
Oxford, 1 Keble Road, Oxford, OX1 3NP, UK}
\author{$^{5}$Ken-Ichi Nakao}
\affiliation{$^{5}$Department of Mathematics and Physics,
Graduate School of Science, Osaka City University,
3-3-138 Sugimoto, Sumiyoshi, Osaka 558-8585, Japan}
\date{\today}
\begin{abstract}
Angular momentum plays very important roles 
in the formation of primordial black holes in the
 matter-dominated phase of the Universe
if it lasts sufficiently long. 
In fact, 
most collapsing masses are bounced back 
due to centrifugal force, since angular momentum 
significantly grows before collapse.
For masses with ${q}\le {q}_{c}\simeq 2.4 {\cal I}^{1/3}\sigma_{H}^{1/3}$, where 
${q}$ is a nondimensional parameter of initial reduced quadrupole moment, 
$\sigma_{H}$ is the density fluctuation at horizon entry $t=t_{H}$, and 
${\cal I}$ is a parameter of the order of unity,
angular momentum gives a suppression factor
$\sim \exp(-0.15 {\cal I}^{4/3} \sigma_{H}^{-2/3})$ to the production
 rate. As for masses with ${q}> {q}_{c}$, 
the suppression factor is even stronger as $\sim 
 \exp(-0.0046{q}^{4}/\sigma_{H}^{2})$.
We derive the spin distribution of primordial black holes
and find that most of the primordial black holes
are rapidly rotating near the extreme value $a_{*}=1$, where 
$a_{*}$ is the nondimensional Kerr parameter at their formation.
The smaller $\sigma_{H}$ is, 
the stronger the tendency towards the extreme rotation.
Combining this result with the effect of anisotropy, 
we numerically and semianalytically estimate the 
production rate $\beta_{0}$ of primordial black holes.
Then we find that
$\beta_{0}\simeq 1.9\times 10^{-7}f_{q}(q_{c}){\cal
       I}^{6}\sigma_{H}^{2}\exp(-0.15{\cal I}^{4/3}\sigma_{H}^{-2/3})$
for $\sigma_{H}\lesssim 0.005$, while $\beta_{0}\simeq
       0.05556\sigma_{H}^{5}$ for $0.005\lesssim \sigma_{H}\lesssim 0.2
       $,
where $f_{q}(q_{c})$ is the fraction of masses whose $q$ is smaller than
       $q_{c}$ and we assume $f_{q}(q_{c})$ is not too small.
We argue that matter domination significantly enhances the production of 
primordial black holes despite the suppression factor.
If the end time $t_{{\rm end}}$ of the matter-dominated phase 
satisfies $t_{{\rm end}}\lesssim (0.4 {\cal I}\sigma_{H})^{-1}t_{H}$, 
the effect of the finite duration significantly suppresses 
primordial black hole formation and weakens the tendency 
towards large spins.

\end{abstract}


\maketitle

\tableofcontents

\newpage

\section{Introduction}
Primordial black holes may have been formed in the early Universe. 
Their masses are given by 
$M\sim (c^{3}/G)t \simeq 10^{15} (t/10^{-23}\mbox{s}) \mbox{g}$, 
where $t$ is the cosmological time of the formation.
They have left observable signatures in the Universe until now.
Observational constraints on the abundance of primordial black holes 
by a variety of observations 
are reviewed in~\cite{Carr:2009jm,Carr:2016drx}.
There still remains a possibility for primordial black holes to be 
a large fraction of dark matter~\cite{Carr:2016drx,Carr:2017jsz}.
Sasaki {\it et al.}~\cite{Sasaki:2016jop} pointed out
that binary primordial black holes 
can be a source of gravitational wave event GW150914 observed
by LIGO~\cite{Abbott:2016blz}. 
See, also,
Refs.~\cite{Bird:2016dcv,Clesse:2016vqa,Raidal:2017mfl} for
other estimates of the merger rate.
This possibility was also discussed for
the recently published event GW170104~\cite{Abbott:2017vtc}.
Pani and Loeb~\cite{Pani:2013hpa} discussed the imprint of
superradiant instabilities of spinning primordial black holes 
on the spectrum of cosmic microwave background. 
Chiba and Yokoyama~\cite{Chiba:2017rvs} 
obtained the spin distribution of
primordial black holes 
and concluded that 
primordial black holes are mostly slowly rotating
based on the critical phenomena in the
collapse of rotating radiation fluid~\cite{Gundlach:2016jzm}.

The primordial black hole formation process was pioneered in the 
radiation-dominated phase of the Universe by Carr~\cite{Carr:1975qj}.
In this phase, there is a threshold $\delta_{{\rm th}}$ of black hole formation, which is 
governed by pressure gradient force, and the production rate of 
black holes is given by $\sim (\delta_{{\rm th}}/\sigma_{H})\exp
[-\delta_{{\rm th}}^{2}/(2\sigma_{H}^{2})]$, where $\sigma_{H}$
is density fluctuation at horizon entry. The threshold $\delta_{{\rm
th}}$ of density perturbation 
was originally estimated to $\sim 1/3$~\cite{Carr:1975qj} and recently
to $\sim 0.42-0.56$ for relatively gentler profiles of 
density field~\cite{Shibata:1999zs,Musco:2004ak,Polnarev:2006aa,Musco:2008hv,Musco:2012au,Harada:2013epa,Harada:2015yda}. On
the other hand, black hole formation in a matter-dominated phase
is not yet studied so much.
A matter-dominated phase is considered 
not only after the matter-radiation equality but also in an  
earlier stage of the Universe, such as the inflaton-oscillating phase after 
inflation~\cite{Assadullahi:2009jc,Alabidi:2009bk,Alabidi:2012ex,Alabidi:2013wtp} 
and the epoch of strong phase 
transition~\cite{Khlopov:1980mg,Polnarev:1982,Sobrinho:2016fay}, 
for which the mass of the 
formed black holes is given in terms of the cosmological 
time of the epoch.

It has been conventionally believed that primordial black holes 
are effectively produced in the matter-dominated phase due to 
the absence or significant reduction 
of the pressure gradient force. The theory of black hole formation 
in the matter-dominated era was pioneered by Khlopov and 
Polnarev~\cite{Khlopov:1980mg,Polnarev:1982}. 
It is deviation from spherical symmetry that governs 
the probability of black hole formation in this phase 
unlike in the radiation-dominated phase. 
In the absence of the pressure gradient force, 
anisotropy develops during collapse so that the final stage can be
described as pancake collapse~\cite{Lin:1965,Zeldovich:1969sb}. 
Harada {\it et al.}~\cite{Harada:2016mhb} reanalyzed this problem and found that 
the application of the hoop conjecture for black hole formation 
results in the production rate
$\beta_{0}\simeq 0.05556 \sigma_{H}^{5}$ for $\sigma_{H}\ll 1$.
Based on this estimate of the production rate, 
Carr {\it et al.}~\cite{Carr:2017edp} discussed the inflaton and spectator 
field perturbations.
It should be noted that the nonspherical effect in primordial black hole
formation was also discussed by K\"uhnel and 
Sandstad~\cite{Kuhnel:2016exn} in a very 
different way.

The effect of rotation has not yet been 
seriously studied in the formation 
of primordial black holes. At first sight, 
it seems negligible because the rotational mode is not growing
in the linear order in cosmological perturbation theory. Even in full
nonlinearity, circulation is conserved in the dynamics of perfect
fluid. In the current paper, however, we show that this expectation is 
not correct. We adopt the theory of angular momentum in structure
formation, which has been developed
to explain the origin of the angular momentum of galaxies by 
Peebles~\cite{Peebles:1969jm} and Catelan and 
Theuns~\cite{Catelan:1996hv}. We find that 
angular momentum plays very important roles in the 
formation of primordial black holes in the matter-dominated phase.

This paper is organized as follows. In Sec.~\ref{sec:preliminaries},
we present basic equations and review cosmological
perturbation theory in Newtonian gravity. 
In Sec.~\ref{sec:angular_momentum}, we introduce the angular momentum of 
masses and review the first-order and second-order contributions.
In Sec.~\ref{sec:application_to_PBH}, we apply this theory to 
primordial black hole formation in the matter-dominated era.
We derive the suppression factors of the production rate
and the initial spin distribution of primordial black holes.
In Sec.~\ref{sec:production_rate_of_PBH}, 
we obtain the production rate of 
primordial black holes in the 
matter-dominated era and discuss it in comparison with 
that in the radiation-dominated era.
Section~\ref{sec:conclusion} is devoted to conclusions.
\section{Preliminaries}
\label{sec:preliminaries}
\subsection{Basic equations}
We briefly review standard cosmological perturbation theory in Newtonian 
gravity. See e.g. 
Peebles~\cite{Peebles:1969jm,Peebles:1994xt} and 
Hwang {\it et al.}~\cite{Hwang:2012bi} for details. We begin with
the Euler equation, the equation of continuity,  
and the Poisson equation:
\begin{eqnarray}
 {\bf a}=-\nabla_{r}\Psi, \quad
 \left(\frac{\partial \rho}{\partial t}\right)_{r}+\nabla_{r}(\rho\cdot
  {\bf v})=0, \quad \mbox{and}\quad 
 \nabla^{2}_{r}\Psi=4\pi G\rho, 
\end{eqnarray}
respectively, where ${\bf v}:={D{\bf r}}/{Dt}$ and ${\bf a}:={D{\bf v}}/{Dt}$,
${\bf r}$ is the Eulerian coordinates, $\nabla_{r}$ 
is the nabla with respect to ${\bf r}$ and $D/Dt$ denotes the 
time derivative along the motion of a fluid element. 
We introduce the comoving coordinates ${\bf x}$, peculiar velocity 
${\bf u}$, density perturbation ${\bf \delta}$, and 
potential perturbation $\psi$ such that
${\bf x}:={\bf r}/{a}$, 
${\bf u}:= a{D{\bf x}}/{Dt}$, 
$\delta:={(\rho-\rho_{0})}/{\rho_{0}}$, and 
$\psi:=\Psi-\Psi_{0}$, 
where $\rho_{0}=\rho_{0}(t)$ and $a=a(t)$ are the density and scale
factor of the homogeneous and isotropic universe, respectively.
Noting
\begin{equation}
 \frac{Df }{Dt}=\frac{\partial f}{\partial
		      t}+\frac{D{\bf x}}{Dt}\cdot
 \nabla f, 
\end{equation}
where $\nabla$ is the nabla with respect to ${\bf x}$,
we find 
\begin{eqnarray}
 {\bf v}=Ha{\bf x}+{\bf u} \quad\mbox{and}\quad  
 {\bf a}=\frac{\partial {\bf u}}{\partial t}
+H{\bf u}+\frac{1}{a}({\bf u}\cdot \nabla){\bf u}+\ddot{a}{\bf x},
\end{eqnarray}
where the dot denotes the derivative with respect to $t$ and 
$
 H:={\dot{a}}/{a}
$
is the Hubble parameter.

As a zeroth-order solution, we find 
\begin{eqnarray}
\rho_{0}a^{3}&=&\mbox{const.} ,
\label{eq:conservation_bg}
\\
 \Psi_{0}&=&\frac{2}{3}\pi G\rho_{0}a^{2}{\bf x}^{2}+C(t), \\
 \frac{\ddot{a}}{a}&=&-\frac{4\pi}{3}G\rho_{0}, \\
 H^{2}&=&\frac{8\pi}{3}G\rho_{0}{\bf -}\frac{K}{a^{2}},
\label{eq:Friedmann}
\end{eqnarray}
where $C(t)$ is an arbitrary function and $K$ is an arbitrary constant.
We assume $K=0$ in this paper, corresponding to the Einstein-de Sitter universe.
Integrating Eq.~(\ref{eq:Friedmann}) 
with Eq.~(\ref{eq:conservation_bg}), we
find 
\begin{equation}
 a(t)=a_{0}t^{2/3},
\label{eq:a(t)}
\end{equation} 
where $a_{0}$ is a positive constant and the integration constant is
chosen so that $a(0)=0$. Equation~(\ref{eq:Friedmann}), hence, yields
\begin{equation}
 \rho_{0}=\frac{1}{6\pi G t^{2}}.
\end{equation}
For the deviation from the zeroth-order solution, we find 
\begin{eqnarray}
 && \frac{\partial {\bf u}}{\partial t}+H{\bf u}+\frac{1}{a}({\bf
  u}\cdot\nabla){\bf u}
  =-\frac{1}{a}\nabla\psi, \label{eq:Euler_eq_Newtonian_cosmology}\\
 && \frac{\partial \delta}{\partial t}+\frac{1}{a}\left[\nabla\cdot {\bf
						   u}+\nabla\cdot
						   (\delta {\bf
						   u})\right]=0, 
\label{eq:continuity_eq_Newtonian_cosmology}
\\
 && \frac{1}{a^{2}}\nabla^{2}\psi=4\pi G\rho_{0}\delta.
\label{eq:Poisson_eq_Newtonian_cosmology}
\end{eqnarray}

\subsection{Linear perturbations}
Linearizing Eqs.~(\ref{eq:Euler_eq_Newtonian_cosmology})
--(\ref{eq:Poisson_eq_Newtonian_cosmology})
and denoting linear perturbations with
\begin{eqnarray}
 {\bf u}_{1}(t, {\bf x}) =\sum_{{\bf k}}\hat{{\bf u}}_{1,{\bf k}}(t)e^{i{\bf k}\cdot
  {\bf x}}, ~
 \delta_{1}(t,{\bf x})=\sum_{{\bf k}}\hat{\delta}_{1,{\bf k}}(t)e^{i{\bf k}\cdot
  {\bf x}}, ~
 \psi_{1}(t,{\bf x})=\sum_{{\bf k}}\hat{\psi}_{1,{\bf k}}(t)e^{i{\bf k}\cdot
  {\bf x}}, 
\end{eqnarray}
we find 
\begin{eqnarray}
 && \dot{\hat{{\bf u}}}_{1, {\bf k}}+H\hat{{\bf u}}_{1, {\bf
  k}}=-i\frac{1}{a}{\bf k}\hat{\psi}_{1, {\bf k}}, \label{eq:u_1k}\\
 && \dot{\hat{\delta}}_{1, {\bf k}}+i\frac{1}{a}{\bf k}\cdot \hat{{\bf
  u}}_{1, {\bf k}}=0, \label{eq:delta_1k}\\
 && -\frac{1}{a^{2}}k^{2}\hat{\psi}_{1,{\bf k}}=4\pi
  G\rho_{0}\hat{\delta}_{1,{\bf k}}.
\label{eq:psi_1k}
\end{eqnarray}
Differentiating Eq.~(\ref{eq:delta_1k}) with respect to $t$
and 
eliminating ${\bf k}\cdot \hat{{\bf u}}_{1,{\bf k}}$ 
and ${\bf k}\cdot \dot{\hat{{\bf u}}}_{1,{\bf
k}}$ by Eqs.~(\ref{eq:u_1k})--
(\ref{eq:psi_1k}), 
we find 
\begin{equation}
 \ddot{\hat{\delta}}_{1, {\bf k}}+\frac{4}{3t}\dot{\hat{\delta}}_{1, {\bf k}}-\frac{2}{3t^{2}}\hat{\delta}_{1, {\bf k}}=0,
\end{equation}
where we have used Eq.~(\ref{eq:a(t)}).
A general solution is given by 
\begin{equation}
 \hat{\delta}_{1, {\bf k}}=A_{{\bf k}}t^{2/3}+B_{{\bf k}}t^{-1},
\end{equation}
where $A_{{\bf k}}$ and $B_{{\bf k}}$ are arbitrary constants. The other
linear perturbations are given by 
\begin{eqnarray}
 \hat{\psi}_{1,{\bf k}}&=& -\frac{2}{3}\frac{a_{0}^{2}}{k^{2}}(A_{{\bf
  k}}+B_{\bf k}t^{-5/3}), \\
 \hat{{\bf u}}_{1,{\bf k}}&=& i a_{0}\frac{{\bf k}}{k^{2}}
\left(\frac{2}{3}A_{{\bf k}}t^{1/3}-B_{{\bf k}}t^{-4/3}\right)+{\bf
C}_{{\bf k}}t^{-2/3}, 
\end{eqnarray}
where ${\bf C}_{{\bf k}}$ is a 
constant vector satisfying ${\bf k}\cdot {\bf C}_{{\bf k}}=0$.
Hereafter, we neglect decaying modes. Then, we find 
\begin{equation}
 {\bf u}_{1}=-\frac{t}{a}\nabla \psi_{1}. 
\label{eq:u1_potential}
\end{equation}
This implies that 
there is a velocity field potential $\phi=(t/a)\psi_{1}$ such that
${\bf u}_{1}=-\nabla \phi$.

\section{Angular momentum}
\label{sec:angular_momentum}

Angular momentum within a comoving region $V$ with respect to the origin of the coordinates is given by 
\begin{eqnarray}
 {\bf L}_{c}=\int_{a^{3}V}\rho {\bf r}\times {\bf v}d^{3}{\bf r} 
=\rho_{0}a^{4}\left(\int_{V}{\bf x}\times {\bf u}d^{3}{\bf x}+\int_{V}{\bf
		 x}\delta\times {\bf u}d^{3}{\bf x}\right).
\label{eq:Lc}
\end{eqnarray}
In Sec.~\ref{sec:2nd_order}, 
we review Peebles's~\cite{Peebles:1969jm} analysis for the second-order contribution 
to the angular momentum.
In Sec.~\ref{sec:1st_order},
we develop a formulation for the first-order contribution 
similar to that Catelan and Theuns~\cite{Catelan:1996hv}  
developed with the Zel'dovich approximation.
\subsection{Second-order contribution}
\label{sec:2nd_order}

If $V$ is a ball centered at the origin, 
the first term in the parentheses on the rightmost side of Eq.~(\ref{eq:Lc}) vanishes to the first order 
because of ${\bf u}_{1}=-\nabla \phi$.
In fact, using Gauss's theorem, we have 
\begin{eqnarray}
 \left[\int_{V}{\bf x}\times \nabla\phi d^{3}{\bf x}\right]_{i}
=\int_{\partial V}\epsilon_{ijk}x_{j}\phi dS_{k},
\end{eqnarray}
which vanishes if $\partial V$ is a sphere.

To see this term beyond the first order, using
Eq.~(\ref{eq:Euler_eq_Newtonian_cosmology}), we obtain
\begin{equation}
 \frac{d}{dt}\left[a\int_{V}{\bf x}\times {\bf u}d^{3}{\bf x}\right]
=-\int_{V}{\bf x}\times ({\bf u}\cdot \nabla ){\bf u}d^{3}{\bf x}
-\int_{V} {\bf x}\times \nabla \psi d^{3}{\bf x}.
\label{eq:time_derivative_first_term}
\end{equation}
Then, the second term on the right-hand side of 
Eq.~(\ref{eq:time_derivative_first_term}) vanishes.
To estimate the first term on the right-hand side of 
Eq.~(\ref{eq:time_derivative_first_term})
to the second order,
we can use the solution ${\bf u}_{1}$ 
of the linear perturbation.
Since
\begin{equation}
 ({\bf u}_{1}\cdot \nabla){\bf u}_{1}=(\nabla \phi \cdot \nabla)\nabla
  \phi=\frac{1}{2}
\nabla (\nabla \phi)^{2},
\end{equation}
the first term on the right-hand side 
of Eq.~(\ref{eq:time_derivative_first_term}) vanishes to the second 
order. Thus, we find the contribution from the 
first term in the parentheses on the rightmost
side of Eq.~(\ref{eq:Lc}) is constant to the second order.

The contribution from the second term in the parentheses 
on the rightmost side of Eq.~(\ref{eq:Lc}) is growing.
We should also note that the center of mass is shifted from the 
origin. The angular momentum ${\bf L}$ with respect to the center of
mass is then given by 
\begin{equation}
 {\bf L}={\bf L}_{c}-{\bf R}\times {\bf P},
\label{eq:L}
\end{equation}
where ${\bf R}$ is the shift of the center of mass and ${\bf P}$
is the linear momentum. We can estimate them to the first order
as
\begin{eqnarray}
&&{\bf R}=\displaystyle\frac{\int_{a^{3}V}\rho {\bf r}d^{3}{\bf r}}{\int_{a^{3}V}\rho
 d^{3}{\bf r}}=  \frac{a}{V}\int_{V} {\bf x}\delta_{1} d^{3}{\bf x}, \\
&&{\bf P}=\int_{a^{3}V}\rho {\bf v}d^{3}{\bf r}=
\rho_{0}a^{3}\int_{V} {\bf u}_{1}d^{3}{\bf
 x}+\rho_{0}a^{3}VH{\bf R}.
\end{eqnarray}

To implement the integration, we use the following formula:
\begin{eqnarray}
 \int_{|{\bf x}|\le r_{0}} e^{i{\bf k}\cdot {\bf x}}d^{3}{\bf
  x}=\frac{4\pi}{3}r_{0}^{3}g(kr_{0}) \quad \mbox{and}\quad 
 \int_{|{\bf x}|\le r_{0}}{\bf x} e^{i{\bf k}\cdot {\bf x}}d^{3}{\bf
  x}=i\frac{4\pi}{15}r_{0}^{5}f(kr_{0}){\bf k},
\label{eq:integral_formula}
\end{eqnarray}
where 
\begin{eqnarray}
 g(y):=3\left(\frac{\sin y}{y^{3}}-\frac{\cos y}{y^2}\right) \quad
  \mbox{and} \quad 
 f(y):=45\left(\frac{\sin y}{y^{5}}-\frac{\cos y}{y^{4}}-\frac{\sin
	  y}{3y^{3}}\right).
\end{eqnarray}
The functions $f$ and $g$ satisfy 
\begin{eqnarray}
 \frac{dg}{dy}=-\frac{y}{5}f, ~~
 \lim_{y\to 0}f(y)=\lim_{y\to 0}g(y)=1, ~~\mbox{and}~~
 \lim_{y\to \infty}f(y)= \lim_{y\to \infty}g(y)=0,
\end{eqnarray}
and show decaying oscillations. They can be regarded
as window functions.

\begin{figure}[htbp]
\begin{center}
 \includegraphics[width=0.6\textwidth]{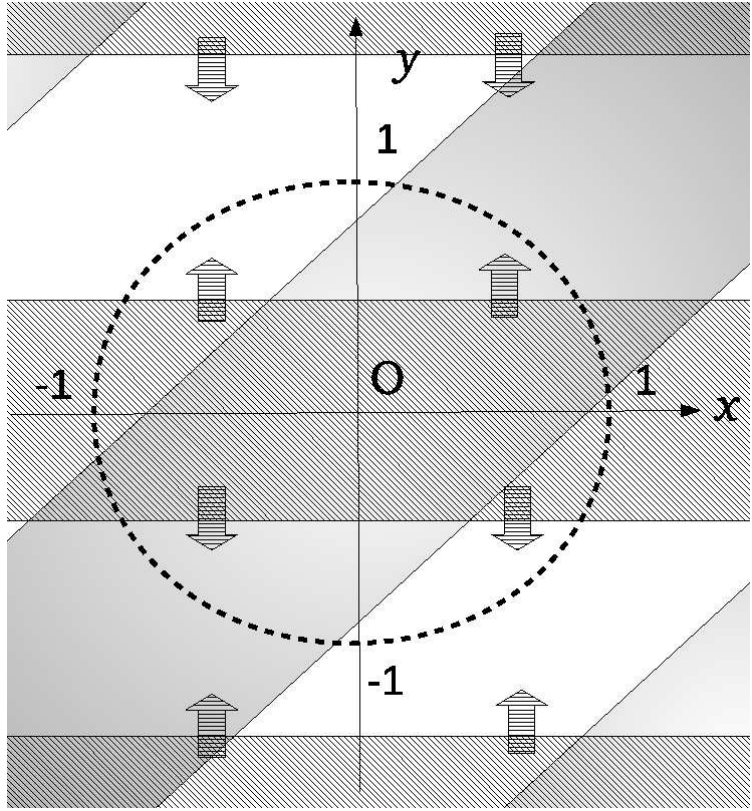} 
\caption{\label{fig:mode_coupling}
The second-order contribution to the angular momentum
comes from the coupling between two independent modes of linear
 perturbations. 
The radius of the ball, $r_{0}$, is chosen to unity.
The regions for $\delta>0$ with ${\bf k}_{1}=(\pi/2,-\pi/2,0)$ are 
shaded. The regions $\psi>0$ with ${\bf k}_{2}=(0,\pi,0)$ are 
hatched and the velocity field is denoted by arrows.
The density field with ${\bf k}_{1}$
and the velocity field with ${\bf k}_{2}$ couple with each
 other so that they compose a growing angular momentum parallel to the $z$ axis
in the anticlockwise direction.}
\end{center}
\end{figure}

Then, we can show the following result:
\begin{eqnarray}
 {\bf R}&=&\frac{i}{5}a t^{2/3}r_{0}^{2}\sum_{{\bf k}}f(kr_{0})A_{{\bf
  k}}{\bf k}, 
\label{eq:R_first}\\
 {\bf P}&=&\frac{8\pi}{9}i\rho_{0}a^{4}t^{-1/3}r_{0}^{3}\sum_{{\bf
 k}}g(kr_{0})A_{{\bf k}}\frac{{\bf k}}{k^{2}}
+\frac{4\pi}{3}\rho_{0}a^{3}r_{0}^{3}H{\bf R}, 
\label{eq:P_first}\\
 {\bf L}_{c}&=& -\frac{8\pi}{45}\rho_{0}(ar_{0})^{5}t^{1/3}\sum_{{\bf
  k},{\bf k}'}A_{{\bf k}}A_{{\bf k}'}\frac{{\bf k}\times {\bf
  k}'}{k^{'2}}
f(|{\bf k}+{\bf k}'|r_{0}).
\label{eq:Lc_second}
\end{eqnarray}
From Eqs.~(\ref{eq:L}) and (\ref{eq:R_first})--(\ref{eq:Lc_second}), we find 
\begin{equation}
 {\bf L}=-\frac{8\pi}{45}\rho_{0}(ar_{0})^{5}t^{1/3}\sum_{{\bf k},{\bf
  k}'}A_{{\bf k}}A_{{\bf k}'}
\frac{{\bf k}\times {\bf k}'}{k^{'2}}\left[f(|{\bf k}+{\bf
				      k}'|r_{0})-f(kr_{0})g(k'r_{0})\right].
\label{eq:L_second_order}
\end{equation}
We can see that ${\bf L}$ increases as $t^{5/3}$ irrespective of the 
details of $A_{{\bf k}}$.
Figure~\ref{fig:mode_coupling} schematically shows that the mode
coupling of two independent modes which are not parallel to each other 
contributes to the growing angular momentum.

Assuming that $A_{{\bf k}}$ takes a random phase, we can calculate the 
variance of ${\bf L}$ as follows:
\begin{eqnarray}
 \langle {\bf L}^{2}\rangle &=&
\left(\frac{8\pi}{45}\rho_{0}(ar_{0})^{5}t^{1/3}\right)^{2}
 \times  \sum_{{\bf k_{1}},{\bf k_{2}}}
\langle |A_{\bf k_{1}}|^{2}\rangle 
\langle |A_{\bf k_{2}}|^{2}\rangle
({\bf k}_{1}\times {\bf k}_{2})^{2} 
\frac{f_{1+2}-f_{1}g_{2}}{k_{2}^{2}} \nonumber \\
&\times & \left[\frac{f_{1+2}-f_{1}g_{2}}{k_{2}^{2}}-\frac{f_{1+2}-f_{2}g_{1}}{k_{1}^{2}}\right],
\label{eq:L_A}
\end{eqnarray}
where $f_{1+2}:=f(|{\bf k}_{1}+{\bf k}_{2}|r_{0})$, $f_{i}:=f(k_{i}r_{0})$ and
$g_{i}:=f(k_{i}r_{0})$ ($i=1,2$).
On the other hand, the density perturbation integrated over 
the ball of radius $r_{0}$ can be calculated to the first order as
\begin{eqnarray}
 \delta_{s}&:=&\frac{\int_{a^{3}V}d^{3}{\bf r}\rho-\int_{a^{3}V}d^{3}{\bf r}\rho_{0}}{\int_{a^{3}V}d^{3}{\bf
  r}\rho_{0}} =\frac{3}{4\pi r_{0}^{3}}\int_{|{\bf x}|<r_{0}}d^{3}{\bf x}\delta
=t^{2/3}\sum_{{\bf k}}A_{\bf k}g(kr_{0}).
\label{eq:delta_s_first_order}
\end{eqnarray}
Thus, the variance of $\delta_{s}$ is given by 
\begin{equation}
 \langle \delta^{2}_{s}  \rangle=t^{4/3}\sum_{{\bf k}}\langle |A_{\bf
  k}|^{2}\rangle g^{2}(k r_{0}).
\label{eq:deltas_A}
\end{equation}
From Eqs.~(\ref{eq:L_A}) and (\ref{eq:deltas_A}), we find 
\begin{equation}
 \langle {\bf L}^{2}\rangle ^{1/2}=
\frac{8\pi}{45}\rho_{0}\frac{(ar_{0})^{5}}{t}{\cal I} \langle
\delta^{2}_{s}\rangle ,
\label{eq:fluctuation_L2}
\end{equation}
where we have defined the ratio ${\cal I}$ as follows:
\begin{equation}
 {\cal I}:=\frac{\left\{\sum_{{\bf k}_{1},{\bf k}_{2}}
\langle |A_{{\bf k}_{1}}|^{2}\rangle 
\langle |A_{{\bf k}_{2}}|^{2}\rangle
({\bf k}_{1}\times {\bf k}_{2})^{2} 
\displaystyle\frac{f_{1+2}-f_{1}g_{2}}{k_{2}^{2}}\left[\frac{f_{1+2}-f_{1}g_{2}}{k_{2}^{2}}-\frac{f_{1+2}-f_{2}g_{1}}{k_{1}^{2}}\right]\right\}^{1/2}}
{\sum_{{\bf k}}\langle |A_{\bf
  k}|^{2}\rangle g^{2}(k r_{0})}.
\end{equation}
Note that ${\cal I}$ does not depend on the overall normalization 
factor of the power spectrum. 
We assume that ${\cal I}$ is of the order of
unity. See~\cite{Peebles:1969jm} for the validity of this assumption.
Equation (\ref{eq:fluctuation_L2}) can be rewritten in the form
\begin{equation}
  \langle {\bf L}^{2}\rangle ^{1/2}=
\frac{2}{15}{\cal I}\frac{MR^{2}}{t} \langle
\delta^{2}_{s}\rangle ,
\label{eq:final_form_second_order}
\end{equation}
where 
$M:=({4\pi}/{3})\rho_{0}(ar_{0})^{3}$ and $R:=a r_{0}$
are the mass and the radius of the ball, respectively.
If $\partial V$ is not a sphere, 
the window functions $f$ and $g$ are 
modified and the factor ${\cal I}$ will be altered. 
However, even in that case, 
Eq.~(\ref{eq:final_form_second_order}) can still apply
as the expression for the second-order contribution.

\subsection{First-order contribution}
\label{sec:1st_order}
If $V$ is not a ball, 
the first term in the parentheses on the rightmost side 
of Eq.~(\ref{eq:Lc}) 
does not vanish in general even in the first order.
To estimate this term, we use Eq.~(\ref{eq:u1_potential}).
We also assume that the Maclaurin-series expansion for $\psi_{1}$
is valid over $V$:
\begin{equation}
 \psi_{1}({\bf x})=\psi_{1}({\bf 0})+\partial_{l}\psi_{1}({\bf
  0})x_{l}+\frac{1}{2}\partial^{2}_{lm}\psi_{1}({\bf 0})x_{l}x_{m}+O(x^{3}),
\end{equation}
where it should be noted that $\psi_{1}$ is time independent.
The truncation of the expansion up to the quadratic terms is 
justified if the wave number ${\bf k}$ of the perturbation 
satisfies $k r_{0}\lesssim  2\pi$, where $r_{0}$ here
stands for the typical size of the region $V$. 
For $k r_{0}\gg 2\pi $, the contribution 
should cancel out after integration over $V$.
Then, we can calculate
\begin{equation}
 -\left[\int_{V}{\bf x}\times \nabla \psi_{1} d^{3}{\bf x}\right]_{i}=\epsilon_{ijk}D_{jm}J_{km},
\end{equation}
where 
\begin{equation}
 J_{jm}:=\int_{V}x_{j}x_{m}d^{3}{\bf x} \quad \mbox{and} \quad
 D_{km}:=\partial^{2}_{km}\psi_{1}({\bf 0}).
\end{equation}
To estimate ${\bf L}$ to the first order, 
we assume that the origin is located at the center of mass 
of $V$.
Noting that only the traceless components of $J_{ij}$ and $D_{ij}$
can contribute to ${\bf L}$, we finally obtain the first-order 
term of ${\bf L}$ as follows:
\begin{equation}
{\bf L}=t\rho_{0}a^{3}\epsilon_{ijk}{\cal D}_{jm}{\cal J}_{km}, 
\end{equation} 
where 
${\cal J}_{ij}:=J_{ij}-\frac{1}{3}\delta_{ij}J_{ll}$ and 
${\cal D}_{ij}:=D_{ij}-\frac{1}{3}\delta_{ij}D_{ll}$.
Note that this contribution grows as $t$.

Assuming that $\partial V$ is determined by an ellipsoid,
which is given by 
\begin{equation}
 \frac{x_{1}^{2}}{A_{1}^{2}}+\frac{x_{2}^{2}}{A_{2}^{2}}+\frac{x_{3}^{2}}{A_{3}^{2}}=1,
\label{eq:ellipsoid}
\end{equation}
the quadrupole
moment $J_{ij}$ of the uniform ellipsoid 
can be easily calculated 
to give
\begin{equation}
 (J_{ij})=\frac{1}{5}V \mbox{diag} (A_{1}^{2},A_{2}^{2},A_{3}^{2})
=\mbox{diag} (i_{1},i_{2},i_{3}), 
\end{equation}
where the coordinate axes are rotated to the major axes and 
$V=(4\pi/3) A_{1}A_{2}A_{3}$. Assuming that 
${\cal J}_{ij}$ and $D_{ij}$ are uncorrelated,  
we obtain 
\begin{equation}
\langle {\bf L}^{2}\rangle =(t\rho_{0}a^{3})^{2}
\epsilon_{ijk}\epsilon_{ipq}
\langle D_{jm}
D_{pl} \rangle 
{\cal J}_{km}
{\cal J}_{ql}.
\label{eq:L^2_linear}
\end{equation} 
We can calculate
\begin{equation}
\langle D_{ij}
D_{kl} \rangle=\frac{4}{9}a_{0}^{4}\sum_{{\bf
k}}\frac{k_{i}k_{j}k_{l}k_{m}}{k^{4}}\langle |A_{{\bf k}}|^{2} \rangle W(kr_{0}), 
\end{equation}
where $W(kr_{0})$ is a window function which satisfies $W(0)=1$ and 
falls off for $kr_{0}\to  \infty$.
If the power spectrum is isotropic, we can find   
\begin{equation}
\sum_{{\bf
k}}\frac{k_{i}k_{j}k_{l}k_{m}}{k^{4}}\langle |A_{{\bf k}}|^{2}
\rangle W(kr_{0})
=\frac{1}{15}
(\delta_{ij}\delta_{lm}+\delta_{il}\delta_{jm}+\delta_{im}\delta_{jl})
 \sum_{{\bf k}}\langle |A_{{\bf k}}|^{2}\rangle W(kr_{0}).
\end{equation}
Using the identity
\begin{equation}
 \epsilon_{ijk}\epsilon_{ipq}(\delta_{jm}\delta_{pl}+\delta_{jp}\delta_{ml}+\delta_{jl}\delta_{mp}){\cal
  J}_{km}{\cal J}_{ql}=3{\cal J}_{ij}{\cal J}_{ij}=2(\mu_{1}^{2}-3\mu_{2}),
\end{equation}
where 
$
 \mu_{1}:=i_{1}+i_{2}+i_{3}
$
and 
$\mu_{2}:=i_{1}i_{2}+i_{2}i_{3}+i_{3}i_{1}$,
Eq.~(\ref{eq:L^2_linear}) is transformed to  
\begin{eqnarray}
 \langle {\bf L}^{2} \rangle &=&(t\rho_{0}a^{3})^{2}\frac{4}{9}a_{0}^{4}
\frac{1}{15}3{\cal J}_{ij}{\cal J}_{ij}\sum_{{\bf k}}\langle |A_{{\bf
k}}|^{2}\rangle W(kr_{0}).
\label{eq:L_first_order}
\end{eqnarray}
Thus, we find   
\begin{eqnarray}
 \langle {\bf L}^{2} \rangle^{1/2} 
 &\simeq & \sqrt{\frac{2}{15}}\frac{2}{3}a_{0}^{2}t \rho_{0}a^{3}
(\mu_{1}^{2}-3\mu_{2})^{1/2}\frac{\langle \delta_{s}^{2}\rangle^{1/2} }
{t^{2/3}},
\label{eq:L1_delta_s'}
\end{eqnarray}
where we have used Eq.~(\ref{eq:deltas_A}) with the 
approximation $W(k r_{0})\simeq g^{2}(kr_{0})$.
We should note $\mu_{1}^{2}-3\mu_{2}\ge 0$, where the equality holds 
if and only if $V$ is an exact ball. 
We can rewrite Eq.~(\ref{eq:L1_delta_s'}) in the following form:
\begin{eqnarray}
 \langle {\bf L}^{2} \rangle^{1/2} 
\simeq \frac{2}{5\sqrt{15}}
{q} \frac{MR^{2}}{t}\langle
\delta_{s}^{2}\rangle^{1/2},
\label{eq:fluctuation_L1}
\end{eqnarray}
where we have chosen $r_{0}=(A_{1}A_{2}A_{3})^{1/3}$ and defined 
\begin{equation}
 {q}:=\sqrt{\frac{{\cal J}_{ij}{\cal J}_{ij}}{3\left(\displaystyle\frac{1}{5}Vr_{0}^{2}\right)^{2}}}=\frac{\sqrt{2(\mu_{1}^{2}-3\mu_{2})}}{\displaystyle\frac{3}{5}V
  r_{0}^{2}}
\end{equation}
as a nondimensional parameter of 
the initial reduced quadrupole moment of the mass. 
In Appendix~\ref{sec:exact_expression_ellipsoid}, we present an exact
expression for the first-order contribution for an ellipsoid without 
invoking the truncated Maclaurin-series expansion and show that 
Eq.~(\ref{eq:fluctuation_L1}) is justified 
if ${q}$ is not too large.

If we can assume that the center of the volume is located at 
the peak of the density field and that $\partial V$ is given by 
an equidensity surface, the distribution of $q$ can be 
inferred by peak theory~\cite{Catelan:1996hv}. 
However, 
we do not need to specify the detailed distribution function of $q$
for the purpose of the current paper.   

\section{Application to primordial black holes}
\label{sec:application_to_PBH}

\subsection{Average angular momentum of masses}
We denote the first-order and second-order contributions, 
which are given by Eqs.~(\ref{eq:fluctuation_L1})
and (\ref{eq:final_form_second_order}), 
with $\langle {\bf L}_{(1)}^{2}\rangle ^{1/2}$
and $\langle {\bf L}_{(2)}^{2} \rangle ^{1/2}$, respectively.
It should be noted that $\langle {\bf L}_{(1)}^{2}\rangle ^{1/2}\propto t$ 
and $\langle {\bf L}_{(2)}^{2}\rangle ^{1/2}\propto t^{5/3}$. 
We can understand these two effects in a unified manner. 
We can see $\langle {\bf L}_{(1)}^{2}\rangle ^{1/2}\propto a q u_{1}\propto
t$, where $q $ is constant in time, while in $\langle {\bf
L}_{(2)}^{2}\rangle ^{1/2}$, the quadrupole moment 
grows as $t^{2/3}$ due to the growth of the density perturbation.
This gives time dependence $t\cdot t^{2/3}=t^{5/3}$ for $\langle
{\bf L}_{(2)}^{2}\rangle^{1/2} $.
Since $\langle {\bf L}^{2} \rangle =\langle {\bf
L}^{2}_{(1)}\rangle +\langle {\bf L}^{2}_{(2)}\rangle $, 
we make an estimate 
$\langle {\bf L}^{2} \rangle^{1/2} \simeq \mbox{max}(\langle {\bf
L}^{2}_{(1)}\rangle^{1/2}, \langle {\bf L}^{2}_{(2)}\rangle^{1/2}) $.

It would be useful to normalize them at the time of horizon entry
$t=t_{H}$, when 
$R=cH^{-1}$. Then, we find
\begin{eqnarray}
 \langle {\bf L}_{(1)}^{2}\rangle ^{1/2}&=&\frac{2}{5\sqrt{15}}{q}
\frac{3GM^{2}}{c}
  \sigma_{H}\left(\frac{t}{t_{H}}\right), \\
 \langle {\bf L}_{(2)}^{2}\rangle ^{1/2}&=&\frac{2}{15}{\cal I}\frac{3GM^{2}}{c}\sigma_{H}^{2}\left(\frac{t}{t_{H}}\right)^{5/3},
\end{eqnarray}
where we have used the relation
$
 {(a(t_{H})r_{0})^{2}}/{t_{H}}={3GM}/{c}
$
and defined 
$
 \sigma_{H}:=\langle \delta_{s,H}^{2} \rangle ^{1/2}$ with $\delta_{s,H}:=\delta_{s}(t_{H})$.
Thus, we can estimate the corresponding nondimensional Kerr parameters
$a_{*}:=L/(GM^{2}/c)$ of the mass:
\begin{eqnarray}
 \langle a_{*(1)}^{2} \rangle ^{1/2}=\frac{2}{5}\sqrt{\frac{3}{5}}{q} 
  \sigma_{H}\left(\frac{t}{t_{H}}\right)~~\mbox{and}~~
 \langle a_{*(2)}^{2}\rangle ^{1/2}=\frac{2}{5}{\cal I}\sigma_{H}^{2}
\left(\frac{t}{t_{H}}\right)^{5/3}.
\end{eqnarray}

To estimate the final value for the angular momentum, 
we take the time of maximum expansion $t_{m}$, when 
nonlinearity becomes important. 
After this time, we can no longer
apply linear perturbation theory. The overdense region 
begins to collapse and separates from the evolution of the 
rest of the Universe. This implies that the angular momentum 
becomes almost constant after $t_{m}$.

The average value of $t_{m}$ can be estimated 
by $\langle \delta_{s}^{2}\rangle^{1/2}=1$ at $t=\langle t_{m} \rangle $. 
We find 
$
 \langle t_{m} \rangle =t_{H}\sigma_{H}^{-3/2}$
from Eq.~(\ref{eq:deltas_A}).
Thus, we can estimate the average value for the Kerr parameter of the
mass as follows:
\begin{eqnarray}
 \langle a_{*}^{2} \rangle ^{1/2}\simeq  \mbox{max}\left(\langle
  a_{*(1)}^{2}\rangle ^{1/2}, \langle a_{*(2)}^{2} \rangle ^{1/2}\right),
\end{eqnarray}
where 
\begin{equation}
 \langle a_{*(1)}^{2}\rangle ^{1/2}=\frac{2}{5}\sqrt{\frac{3}{5}}{q}  \sigma_{H}^{-1/2}~~\mbox{and}~~
 \langle a_{*(2)}^{2}\rangle ^{1/2}=\frac{2}{5}{\cal I}\sigma_{H}^{-1/2}.
\end{equation}
If $q =O(1)$, we find that the first-order effect 
is comparable with the second-order effect.
If we assume $\sigma_{H}\lesssim 0.1$, we have 
$\langle a_{*}^{2}\rangle ^{1/2}\gtrsim 1$, 
implying that centrifugal force 
will prevent the direct collapse to a black hole. Only the 
masses satisfying $a_{*}\le 1$, which are the minority, 
can directly collapse to a black hole.
Therefore, primordial black hole formation
is strongly suppressed by centrifugal force. 
Most of the primordial black holes are rapidly rotating at least when they 
are formed.
The above argument must be weakened if 
the matter-dominated era does not last sufficiently long.
This possibility will be discussed later.

\subsection{Hypothesis}
Although the above discussion qualitatively indicates 
the crucial role of angular 
momentum and the rapid rotation of black holes, it does not tell how the Kerr parameter of the mass is
distributed because we have only estimated the average value of 
$t_{m}$.

To circumvent the difficulty in determining the distributional 
properties of the Kerr parameter, 
we make an assumption.
In Eqs.~(\ref{eq:L_second_order}), (\ref{eq:delta_s_first_order}), 
and (\ref{eq:L_first_order}),
we can see that $\langle {\bf L}_{(1)}^{2}\rangle $, $\langle {\bf
L}_{(2)}^{2}\rangle $,  
and $\langle \delta_{s}^{2}\rangle $ consist of 
the coupling of modes.
The crucial difference is that $\langle {\bf L}_{(2)}^{2} \rangle $
consists of the mode coupling of two independent 
modes which are not parallel to
each other, while both $\langle {\bf L}_{(1)}^{2}\rangle $ and 
$\langle \delta_{s}^{2}\rangle $ consist of the self-coupling of 
a single mode. 

Therefore, despite the complicated dependence of ${\bf L}$ on $A_{{\bf k}}$, 
it is natural to assume that 
$|{\bf L}_{(1)}|\propto \delta_{s}$ and
$|{\bf L}_{(2)}|\propto \langle \delta^{2}_{s} \rangle ^{1/2}
\delta_{s}$,
where and hereafter we focus on the overdense regions.
More precisely, 
inspired by Eqs.~(\ref{eq:fluctuation_L1}) and 
(\ref{eq:final_form_second_order}), we adopt
the following simple approximation 
\begin{eqnarray}
|{\bf L}_{(1)}|&\simeq &\frac{2}{5\sqrt{15}} {q}  \frac{MR^{2}}{t}
\delta_{s}, 
\label{eq:L1_delta_s}
\\
 |{\bf L}_{(2)}|&\simeq& 
\frac{2}{15}{\cal I}\frac{MR^{2}}{t}
\langle \delta^{2}_{s} \rangle ^{1/2} \delta_{s}.
\label{eq:L2_delta_s^2}
\end{eqnarray}
This can be tested by the Monte Carlo simulation.
This is consistent with 
Eqs.~(\ref{eq:fluctuation_L1}) and (\ref{eq:final_form_second_order}).
This also implies 
that the angular momentum is larger for the mass
with larger density perturbation at the same time. 
Here we explain the motivation of Eq.~(\ref{eq:L2_delta_s^2}).
At a density peak, it is most probable that 
a single mode is excited to a large amplitude, while 
others are kept to average ones.
Note that $\delta_{s}$ can be excited by a 
single mode, while ${\bf L}_{(2)}$ only by the 
coupling of two independent modes. So, we can estimate $|{\bf L}_{(2)}|\propto
\langle \delta_{s}^{2}\rangle ^{1/2} \delta_{s}$.
This assumption will be justified particularly if 
$\delta_{s}\gg \langle \delta_{s}^{2}\rangle ^{1/2}$.
On the other hand, 
the above approximation is not correct 
if the wave numbers of all nontrivial 
modes are parallel to each other, where 
${\bf L}_{(2)}=0$ but $\delta_{s}\ne 0$. 
With such an exceptional case, we
assume that Eqs.~(\ref{eq:L1_delta_s}) and (\ref{eq:L2_delta_s^2}) 
are valid for almost all cases.

We estimate $t_{m}=t_{H}(\delta_{s,H})^{-3/2}$ from
Eq. (\ref{eq:delta_s_first_order}).
Then, from Eqs. (\ref{eq:L1_delta_s})
and (\ref{eq:L2_delta_s^2}), we can estimate $a_{*}$ 
as 
\begin{eqnarray}
a_{*}\simeq \mbox{max}(a_{*(1)},a_{*(2)}),
\end{eqnarray}
where
\begin{eqnarray}
 a_{*(1)}=\frac{2}{5}\sqrt{\frac{3}{5}}{q} 
  (\delta_{s,H})^{-1/2}~\mbox{and}~
 a_{*(2)}=\frac{2}{5}{\cal I}\sigma_{H}(\delta_{s,H})^{-3/2}.
\label{eq:a*1_a*2}
\end{eqnarray}
The dependence on $\delta_{s,H}$ can be understood as follows.
Since $t_{m}$ is proportional to $(\delta_{s,H})^{-3/2}$,
it takes a longer time for the mass with smaller $\delta_{s,H}$
to get into the nonlinear regime. This longer 
$t_{m}$ gives a longer time for the angular momentum to grow and 
this growth overcompensates the smaller initial value for the seed angular
momentum. That is, the smaller $\delta_{s,H}$ is, the larger the
final value for $a_{*}$ becomes. This is the case both for the first-order and 
second-order contributions.

Of course, it should be noted that if $\delta_{s,H}$
is too small, $t_{m}$ can be later
than $t_{{\rm end}}$, the end time of the matter-dominated era.
The finite duration of the matter-dominated era, thus, will 
give a lower cutoff 
\begin{equation}
\delta_{{\rm fd}}:=(t_{H}/t_{{\rm end}})^{2/3}
\end{equation}
on $\delta_{s,H}$ below which no primordial black hole is formed.

\subsection{Suppression to primordial black hole production}
We can find that the equality $a_{*(1)}=a_{*(2)}$ holds for
$\delta_{s,H}=\delta_{s,Ht}$, where
$
 \delta_{s,Ht}:=
\sqrt{{5}/{3}}
{\cal  I}{q} ^{-1}\sigma_{H}$.
This determines a transition point between the 
two cases $a_{*}\simeq a_{*(1)}>a_{*(2)}$
and $a_{*}\simeq a_{*(2)}>a_{*(1)}$.
Thus, if $\delta_{s,H}<\delta_{s,Ht}$ for which $a_{*t}\le a_{*}$, 
$a_{*}\simeq a_{*(2)}> a_{*(1)}$, 
while if $\delta_{s,H}>\delta_{s,Ht}$ for which
$0\le a_{*}<a_{*t}$, $a_{*}\simeq a_{*(1)}>a_{*(2)}$, where
$
 a_{*t}:=(2/5)({3}/{5})^{3/4}{\cal I}^{-1/2}{q}
 ^{3/2}\sigma_{H}^{-1/2}$.

We adopt the Kerr bound $a_{*}\le 1$ as the condition for the direct formation of a
black hole. From Eq.~(\ref{eq:a*1_a*2}), we can find
that this condition reduces to $\delta_{s,H}\ge \delta_{{\rm th}}$. 
The threshold $\delta_{{\rm th}}$ is given by 
\begin{eqnarray}
 \delta_{{\rm th}}= \mbox{max}(\delta_{{\rm th}(1)},\delta_{{\rm
  th}(2)},\delta_{{\rm fd}}),
\end{eqnarray}
where 
\begin{eqnarray}
\delta_{{\rm th}(1)}:=\frac{3\cdot 2^{2}}{5^{3}}{q}^{2}~~
\mbox{and}~~
\delta_{{\rm th}(2)}:=\left(\frac{2}{5}{\cal I}\sigma_{H}\right)^{2/3}.
\label{eq:threshold}
\end{eqnarray}
If $\delta_{{\rm fd}}<\delta_{{\rm th}(2)}$ or 
\begin{equation}
 t_{{\rm end}}>\left(\frac{2}{5}{\cal I}\sigma_{H}\right)^{-1}t_{H},
\label{eq:long_MD}
\end{equation}
we can neglect the effect of finite duration. Otherwise, 
the primordial black hole formation is significantly suppressed and 
the tendency towards large spins is weakened. 
This effect is very sensitive to the cosmological scenario. 
In Appendix~\ref{sec:reheating_temperature}, we briefly discuss 
this effect in terms of the reheating temperature.
Here we focus on the effect of angular momentum 
in the following analysis simply by assuming Eq.~(\ref{eq:long_MD}).
We find that $\delta_{{\rm th}(1)}=\delta_{{\rm th}(2)}$ if and only if ${q}={q}_{c}$,
where 
\begin{equation}
 {q}_{c}=\sqrt{\frac{2}{3}}\left(\frac{5}{2}\right)^{7/6}{\cal
  I}^{1/3}\sigma_{H}^{1/3}.
\label{eq:qc}
\end{equation}
Note that $a_{*t}$ can be
rewritten in the form $a_{*t}=({q}/{q}_{c})^{3/2}$ in terms
of 
${q}$ and
${q}_{c}$. 
If ${q}>{q}_{c}$, $a_{*t}>1$, while if ${q}<{q}_{c}$, $a_{*t}<1$.

\begin{figure}[htbp]
\begin{center}
 \includegraphics[width=0.6\textwidth]{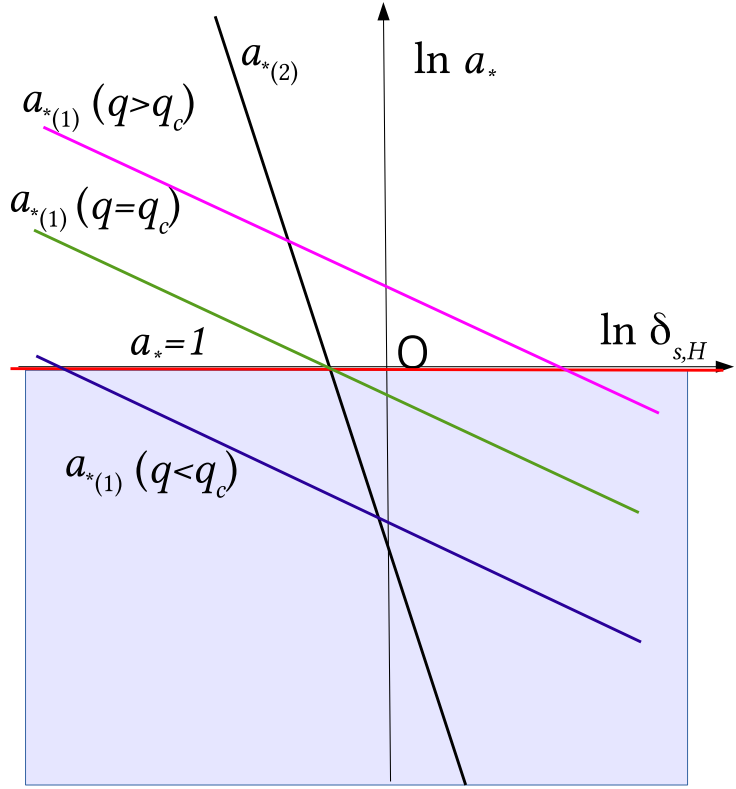} 
\caption{\label{fig:a_delta}
The first-order and second-order contributions to the Kerr parameter, 
$a_{*(1)}$ and $a_{*(2)}$, are schematically plotted as functions of 
the density perturbation at horizon entry, $\delta_{s,H}$.
$a_{*(1)}(\delta_{s,H})$ is plotted for three cases,
 ${q}>{q}_{c}$, ${q}={q}_{c}$, and 
${q}<{q}_{c}$.
The Kerr parameter $a_{*}$ is determined by $a_{*}\simeq
 \mbox{max}(a_{*(1)},a_{*(2)})$. The intersection of
 $a_{*(1)}(\delta_{s,H})$ and $a_{*(2)}(\delta_{s,H})$
 corresponds to the transition point
 $(\delta_{s,H},a_{*})=(\delta_{s,Ht},a_{*t})$. We can see 
$a_{*t}>1$, $a_{*t}=1$, and $a_{*t}<1$ for ${q}>{q}_{c}$, 
${q}={q}_{c}$, 
and ${q}<{q}_{c}$, respectively. 
The horizontal line $a_{*}=1$, which is 
denoted with a red solid line, corresponds to the extreme spin. 
The region below $a_{*}=1$, which is colored in pale blue, 
is that for black hole formation satisfying the Kerr bound $a_{*}\le 1$.
We can see that the threshold $\delta_{{\rm th}}$ is determined by
 $a_{*(1)}$ and rapidly increasing with respect to $q$ 
for ${q}>{q}_{c}$, while it is determined by $a_{*(2)}$ and constant 
for ${q}\le {q}_{c}$.
}
\end{center}
\end{figure}

Figure~\ref{fig:a_delta} schematically 
shows how the Kerr bound $a_{*}\le 1$
gives the threshold $\delta_{\rm th}$ 
for $\delta_{s,H}$ depending on the value of ${q}$, 
where $a_{*}\simeq \mbox{max}(a_{*(1)},a_{*(2)})$. 
If ${q}>{q}_{c}$, then 
$\delta_{{\rm th}}=\delta_{{\rm th}(1)}>\delta_{{\rm th}(2)}$, while 
if ${q}<{q}_{c}$, 
$\delta_{{\rm th}}=\delta_{{\rm th}(1)}<\delta_{{\rm th}(2)}$.

It is natural to assume that ${q}_{c}(\simeq \sigma_{H}^{1/3})$ is small, 
while there is no {\it a priori}
reason for ${q}  $ to be perturbatively small. 
If ${q}> {q}_{c}$, 
we can adopt 
$
 \delta_{{\rm th}}=\delta_{{\rm th}(1)}=({3\cdot 2^{2}}/{5^{3}}){q}^{2}$.
This implies an exponential suppression factor to black
hole formation probability. 
We use the following formula for a Gaussian distribution:
\begin{equation}
P= 2\int_{\delta_{{\rm th}}}^{\infty}d\delta \frac{1}{\sqrt{2\pi
  \sigma^{2}}}\exp\left(-\frac{\delta^{2}}{2\sigma^{2}}\right)
=\mbox{erfc}\left(\frac{\delta_{{\rm th}}}{\sqrt{2}\sigma}\right)
\simeq 
\sqrt{\frac{2}{\pi}}\frac{\sigma}{\delta_{{\rm th}}}
\exp\left(-\frac{\delta_{{\rm th}}^{2}}{2\sigma^{2}}\right), 
\label{eq:exponential_suppression}
\end{equation}
where in the last equality we have assumed $\delta_{{\rm th}}\gg \sigma$
and used an approximation
$
 \mbox{erfc}(x)\simeq {e^{-x^{2}}}/({x\sqrt{\pi}})
$
for $x\gg 1$. 
This results in the following suppression factor
\begin{equation}
 P_{{\rm am}(1)}= \mbox{erfc}\left(\frac{1}{\sqrt{2}\sigma_{H}}
\frac{3\cdot 2^{2}}{5^{3}}{q}^{2}\right)\simeq 
\sqrt{\frac{2}{\pi}}\left(\frac{3\cdot 2^{2}}{5^{3}}\right)^{-1}{q}^{-2}
\sigma_{H}
\exp\left[-\left(\frac{3\cdot
				     2^{2}}{5^{3}}\right)^{2}
\frac{{q}^{4}}{2\sigma_{H}^{2}}\right].
\end{equation}
In the above, we can see that primordial black holes are dominated by
masses with smaller ${q}$. This motivates us to see masses with
${q}< {q}_{c}$, which are the minority of all masses.

For ${q}< {q}_{c}$, 
the threshold $\delta_{\rm th}$ is
given by 
$
 \delta_{{\rm th}}=\delta_{{\rm th}(2)}=\left[({2}/{5}){\cal I}\sigma_{H}\right]^{2/3}$.
While this threshold depends on $\sigma_{H}$, 
$\delta_{{\rm th}}\gg \sigma_{H}$ is still satisfied if $\sigma_{H}\ll 1$.
We can estimate a suppression factor through
Eq.~(\ref{eq:exponential_suppression}) as follows:
\begin{eqnarray}
P_{{\rm am}(2)}=\mbox{erfc}\left[\frac{1}{\sqrt{2}}\left(\frac{2}{5}{\cal I}\right)^{2/3}\sigma_{H}^{-1/3}\right]
\simeq \sqrt{\frac{2}{\pi}}\left(\frac{2}{5}{\cal
			I}\right)^{-2/3}\sigma_{H}^{1/3}
\exp \left[-\frac{1}{2}\left(\frac{2}{5}{\cal
			I}\right)^{4/3}\sigma_{H}^{-2/3}\right].
\end{eqnarray}
Since this suppression is much weaker than that from the first-order
effect if $\sigma_{H}\ll 1$, 
we can conclude that primordial black holes are dominated 
by the masses with ${q}< {q}_{c}$ if ${q}$ is distributed 
around $0$, although this suppression factor is still exponential.

\subsection{Distribution of spins of primordial black holes}
Since
\begin{eqnarray}
 a_{*(1)}^{-2}=\left[\frac{2}{5}\sqrt{\frac{3}{5}}{q} \right]^{-2} \delta_{s,H} ~~\mbox{and}~~
 a_{*(2)}^{-2/3}=\left(\frac{2}{5}{\cal I}\sigma_{H}\right)^{-2/3}\delta_{s,H},
\label{eq:a*1^(-2)_a*2^(-2/3)}
\end{eqnarray}
$a_{*(1)}^{-2}$ and $a_{*(2)}^{-2/3}$ obey Gaussian
distributions centered at 0 with standard deviations
\begin{eqnarray}
 \sigma_{a_{*(1)}^{-2}}=\left[\frac{2}{5}\sqrt{\frac{3}{5}}{q} \right]^{-2}
  \sigma_{H}~~\mbox{and}~~
 \sigma_{a_{*(2)}^{-2/3}}=\left(\frac{2}{5}{\cal I}\right)^{-2/3}\sigma_{H}^{1/3},
\end{eqnarray}
respectively, if they are appropriately extended to the whole real axis.

\begin{figure}[htbp]
 \begin{center}
 \includegraphics[width=0.6\textwidth]{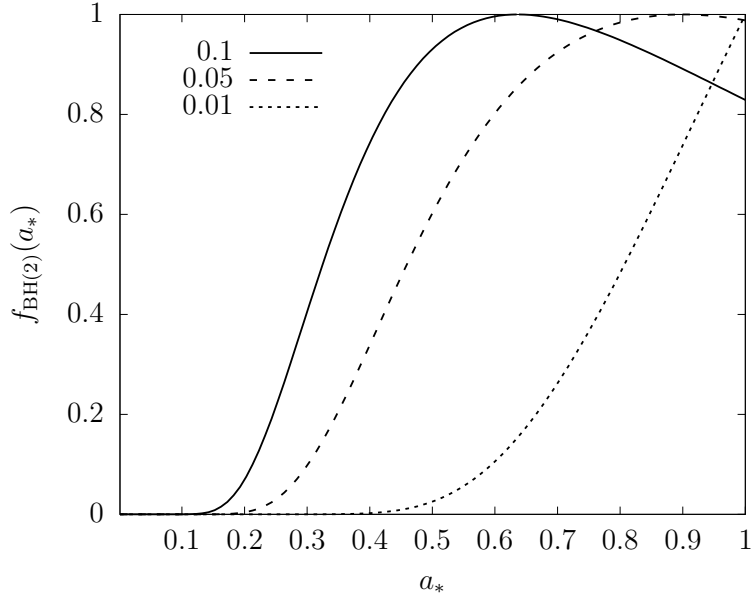}
\caption{\label{fig:spin_2nd}
The spin distribution of primordial black holes formed in the
  matter-dominated era due to the second-order effect, which 
applies for $a_{*t}<a_{*}$, where $a_{*t}=({q}/{q}_{c})^{3/2}$.
We put ${\cal I}=1$. The curves denote the 
spin distribution functions normalized by 
their maximum values for density fluctuations
  $\sigma_{H}=0.1$, $0.05$, and $0.01$. 
We can see that 
 the distribution has a peak at $a_{*}\simeq 0.63$ for
  $\sigma_{H}=0.1$. 
The peak value for $a_{*}$
increases as $\sigma_{H}$ is decreased and reaches the extreme
value $a_{*}=1$ for $\sigma_{H}\simeq 0.04$. The peak lies at $a_{*}=1$ for
  $\sigma_{H}\lesssim 0.04$. It becomes sharper and sharper as
  $\sigma_{H}$ is decreased further.
It should be noted that we 
have neglected possible change in the spin due to 
  the general relativistic dynamics of the formation process as well 
  as mass accretion and quantum radiation after formation.}
 \end{center}
\end{figure}

As we have seen, masses with $q<q_{c}$ dominate
primordial black holes if $q$ is distributed around $0$. 
In this case, $a_{*}\simeq a_{*(2)}\ge a_{*(1)}$ for $a_{*t}\le a_{*}\le
1$, while 
$a_{*}\simeq a_{*(1)}\ge a_{*(2)}$ for $0\le a_{*} <a_{*t}$.
For $a_{*t}\le a_{*}\le 1$,  
since $a_{*}^{-2/3}$ obeys a Gaussian distribution, 
we can estimate the probability $f_{{\rm BH}}(a_{*})da_{*}$ 
for $a_{*}$ of the black hole to be between
$a_{*}$ and $a_{*}+da_{*}$ as
\begin{eqnarray}
 f_{{{\rm BH}}(2)}(a_{*})da_{*} \propto 
\frac{1}{a_{*}^{5/3}}\exp\left(-\frac{1}{2\sigma_{H}^{2/3}}
\left(\frac{2}{5}{\cal I}\right)^{4/3}\frac{1}
{a_{*}^{4/3}}\right)da_{*}.
\end{eqnarray}
For $0\le a_{*} < a_{*t}$, 
since $a_{*}^{-2}$ obeys a Gaussian distribution,
we can estimate $f_{{\rm BH}}(a_{*})$ as 
\begin{eqnarray}
 f_{{\rm BH}(1)}(a_{*})da_{*} &\propto &
\frac{1}{a_{*}^{3}}\exp\left(-\frac{1}{2\sigma_{H}^{2}}\frac{3^{2}2^{4}}{5^{6}}\frac{{q}^{4}}{a_{*}^{4}}\right)da_{*}.
\end{eqnarray}
Since the distribution function is continuous at 
$a_{*}=a_{*t}$, we find that $f_{{\rm BH}}(a_{*})$ is given by 
\begin{eqnarray}
f_{{\rm BH}}(a_{*})=\left\{\begin{array}{cc}
          f_{{\rm BH}(1)}(a_{*})\displaystyle\frac{f_{{\rm
	   BH}(2)}(a_{*t})}{f_{{\rm BH}(1)}(a_{*t})} & (0\le a_{*} < a_{*t})\\
	  f_{{\rm BH}(2)}(a_{*})& (a_{*t}\le a_{*}\le 1) 
		\end{array}
\right.
\end{eqnarray}
up to the overall normalization factor.
Figure~\ref{fig:spin_2nd} shows the distribution of the 
black hole spins due to the second-order effect $f_{{\rm
BH}(2)}(a_{*})$,
where we have chosen ${\cal I}=1$.
We can see that most of the black holes are rapidly rotating.
For $\sigma_{H}=0.1$, the most frequent value for the spin
is given by $a_{*}\simeq 0.63$.
If $\sigma_{H}\gtrsim 0.04$, the most frequent value 
is smaller than the extreme value $a_{*}=1$, while 
it becomes the extreme value for $\sigma_{H}\lesssim 0.04$.
The distribution becomes sharper and sharper at
$a_{*}=1$ as the density fluctuation $\sigma_{H}$ is decreased further.
The black hole with $a_{*}\lesssim 0.2$ is very rare for a reasonable range
of $\sigma_{H}$. 
For clarity, we do not plot the switch to the first-order
effect for $0\le a_{*} <a_{*t}$ in this figure.
In fact, the switching for $0\le a_{*}<a_{*t}$
does not change the qualitative
behavior of the spin distribution function very much.
It should be noted that we neglect possible 
change in the spin due to  
  the general relativistic dynamics of the formation process as well 
  as mass accretion and quantum radiation after formation.

It should be noted that the third-order and higher-order contributions
can be as large as the second-order one at the maximum expansion. 
Generally speaking, higher-order effects will add more variance 
to the Kerr parameter. 
This suggests that the current analysis up to the second-order 
contribution can be valid in order of magnitude,
although the higher-order contributions are yet to be studied.

\begin{figure}[htbp]
 \begin{center}
 \includegraphics[width=0.6\textwidth]{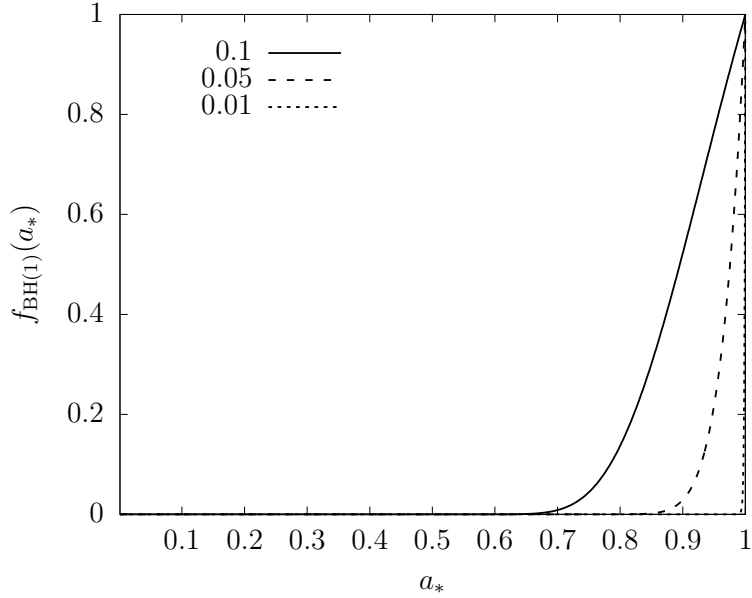}
\caption{\label{fig:spin_1st}
The spin distribution of primordial black holes formed in the
  matter-dominated era due to the first-order effect, 
which applies for $0\le a_{*}<a_{*t}$, where 
$a_{*t}=({q}/{q}_{c})^{3/2}$. 
We choose the quadrupole parameter ${q}$ to $\sqrt{2}$. 
The curves denote the spin distribution functions normalized by 
their maximum values for density fluctuations
  $\sigma_{H}=0.1$, $0.05$, and $0.01$. 
We can see that 
 the distribution has a peak at $a_{*}=1$ and it becomes sharper and
  sharper as $\sigma_{H}$ is decreased. It should be noted that we 
have neglected possible change in the spin due to  
  the general relativistic dynamics of the formation process as well 
  as mass accretion and quantum radiation after formation.}
 \end{center}
\end{figure}

It is also interesting to see black hole spin distribution 
for ${q}> {q}_{c}$, where $a_{*}\simeq a_{*(1)}> a_{*(2)}$
for $0\le a_{*}\le 1$. 
In this regime, we find $f_{{\rm BH}}(a_{*})=f_{{\rm BH}(1)}(a_{*})$
up to the overall normalization.
Figure~\ref{fig:spin_1st} shows the distribution of the 
black hole spin due to the first-order effect, $f_{{\rm BH}(1)}(a_{*})$. 
The exponential dependence indicates that 
the spin parameter distribution is very dense near $a_{*}= 1$, while
it is extremely sparse for $a_{*}\lesssim 0.6$.
We can see that the tendency towards the extreme 
rotation is much stronger than that for $f_{{\rm BH}(2)}(a_{*})$

\section{Production rate}
\label{sec:production_rate_of_PBH}

\subsection{Production rate in the matter-dominated era}
It would be interesting to calculate the probability of 
black hole formation by combining the effects of 
angular momentum and anisotropic collapse, the latter of which has been 
studied by Khlopov and Polnarev~\cite{Khlopov:1980mg,Polnarev:1982} 
and refined by Harada {\it et al.}~\cite{Harada:2016mhb}.

To proceed further, we briefly introduce the Zel'dovich approximation, 
where the location of the fluid element is given by 
\begin{equation}
 {\bf r}=a(t){\bf q}+b(t){\bf p}({\bf q}),
\label{eq:ZA}
\end{equation}
where ${\bf q}$ is the Lagrangian coordinates and 
$b(t)$ is a growing mode of linear perturbation in Newtonian gravity.
We introduce the eigenvalues $\alpha$, $\beta$, and $\gamma$ of the 
tensor $-\partial p_{i}/\partial q_{j}$ and assume $\alpha\ge \beta\ge \gamma$ 
without loss of generality. Taking the normalization 
$b(t_{H})=a(t_{H})$, the linear density perturbation at the horizon entry 
can be given by 
$
 \delta_{s,H}(\alpha,\beta,\gamma)=\alpha +\beta+\gamma.
$
The Zel'dovich approximation~\cite{Zeldovich:1969sb} is the extrapolation 
of Eq.~(\ref{eq:ZA}) beyond the linear regime. 
The probability distribution function for $\alpha$, $\beta$, and $\gamma$
is given by Doroshkevich~\cite{Doroshkevich:1970} as 
\begin{eqnarray}
 w(\alpha,\beta,\gamma)&=&-\frac{27}{8\sqrt{5}\pi \sigma_{3}^{6}}
\exp\left[-\frac{3}{5\sigma_{3}^{2}}\left\{(\alpha^{2}+\beta^{2}+\gamma^{2})-\frac{1}{2}(\alpha\beta+\beta\gamma+\gamma\alpha)\right\}\right]
\nonumber \\
&& \cdot (\alpha-\beta)(\beta-\gamma)(\gamma-\alpha)d\alpha d\beta
 d\gamma,
\label{eq:Doroshkevich}
\end{eqnarray}
where we can find the relation $\sigma_{H}=\sqrt{5}\sigma_{3}$.

Assuming that a mass to be a black hole is initially given by a ball,
the hoop conjecture for black hole formation applied to pancake collapse
implies $h(\alpha,\beta,\gamma)\lesssim 1$, where
\begin{equation}
 h(\alpha,\beta,\gamma):=\frac{2}{\pi}\frac{\alpha-\gamma}{\alpha^{2}}E
\left(\sqrt{1-\left(\frac{\alpha-\beta}{\alpha-\gamma}\right)^{2}}\right)
\end{equation}
and $E(e)$ is the complete elliptic integral of the 
second kind~\cite{Harada:2016mhb}.
The production rate $P_{{\rm ai}}$ due to this effect is calculated by 
\begin{equation}
 P_{{\rm ai}}\simeq \int_{0}^{\infty}d\alpha \int_{-\infty}^{\alpha}d\beta
  \int_{-\infty}^{\beta}d\gamma
  \theta[1-h(\alpha,\beta,\gamma)]w(\alpha,\beta,\gamma).
\label{eq:triple_integral_anisotropy}
\end{equation}
Harada {\it et al.} numerically 
calculated this integral and plotted the result 
in Fig.~1 in~\cite{Harada:2016mhb}.
They also obtain a semianalytic formula
\begin{equation}
P_{{\rm ai}} \simeq 0.05556\sigma_{H}^{5}.
\label{eq:semianalytic_anisotropy}
\end{equation}

\begin{figure}[htbp]
 \begin{center}
   \includegraphics[width=0.7\textwidth]{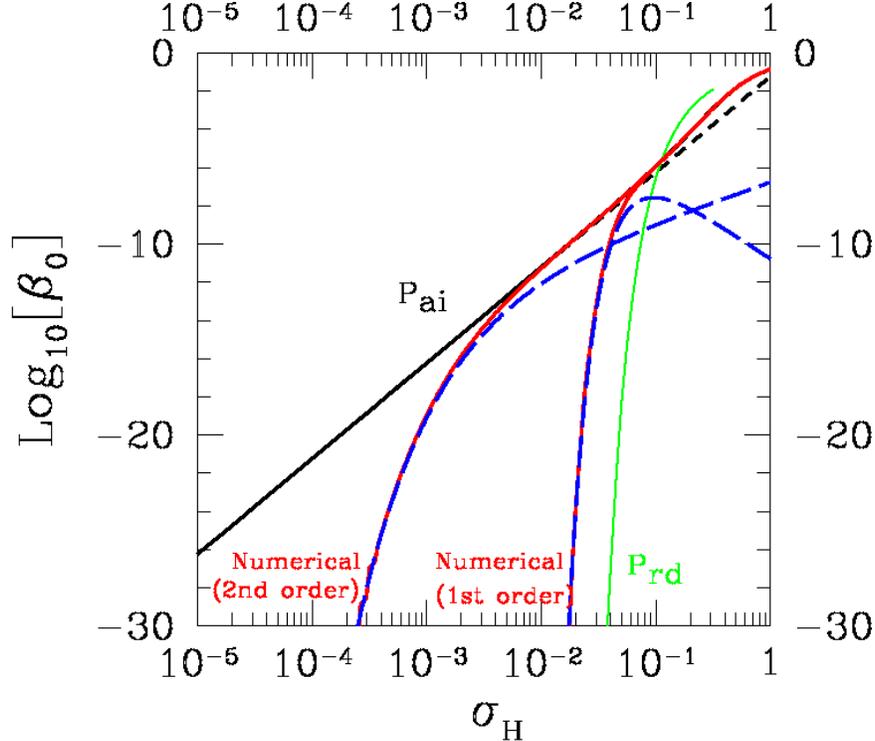}
\caption{\label{fig:production_rate}
The production rates of primordial black holes are plotted. 
We plot the results of numerical integration for $P_{ai}$ due to the 
effect of anisotropic
  collapse, $\beta_{0(2)}$ due to the combined effects of anisotropic 
collapse and second-order angular momentum, 
and $\beta_{0(1)}$ due to the combined effects of anisotropic 
collapse and first-order angular momentum with 
a black solid line labeled $P_{{\rm ai}}$,
a red solid line labeled ``(2nd order)'', and 
a red solid line labeled ``(1st order)'', respectively.
We also plot the corresponding semianalytic formulas
with a black short dashed line labeled $P_{{\rm ai}}$ and 
blue long dashed lines labeled ``(2nd order)'' and ``(1st order)'', 
respectively.
For comparison, we also plot the production rate in the 
radiation-dominated era with a green solid line labeled $P_{\rm rd}$. 
For $\beta_{0(2)}$, $\beta_{0(1)}$, and $P_{\rm rd}$, 
we choose ${\cal I}=1$, 
$q=\sqrt{2}$, and $\delta_{{\rm th}}=0.42$, respectively. 
If $q$ is distributed around $0$, $\beta_{0}\simeq f_{q}(q_{c})\beta_{0(2)}$
applies for $\sigma_{H}\lesssim
  0.005$, where $f_{q}(q_{c})$ denotes the fraction of masses of which $q$ is
  smaller than $q_{c}$, 
while $\beta_{0}\simeq P_{{\rm ai}}$ applies for $0.005 \lesssim \sigma_{H} \lesssim  1$.}
 \end{center}
\end{figure}

In the current paper, we have found that the threshold 
for black hole formation $\delta_{{\rm th}}$ due to the effect of
angular momentum is given by Eq.~(\ref{eq:threshold}). Thus, the production rate of primordial 
black holes can be calculated by 
\begin{equation}
 \beta_{0}\simeq \int_{0}^{\infty}d\alpha \int_{-\infty}^{\alpha}d\beta
  \int_{-\infty}^{\beta}d\gamma
  \theta[\delta_{H}(\alpha,\beta,\gamma)-\delta_{{\rm
  th}}]\theta[1-h(\alpha,\beta,\gamma)]w(\alpha,\beta,\gamma).
\label{eq:triple_integral}
\end{equation}
To see the second-order and
first-order effects separately, we put $\delta_{{\rm th}}=\delta_{{\rm
th}(2)}$ and $\delta_{{\rm th}}=\delta_{{\rm
th}(1)}$ in Eq.~(\ref{eq:semianalytic_anisotropy_threshold}) and 
denote them with $\beta_{0(2)}$ and $\beta_{0(1)}$, respectively.
We have numerically implemented triple integration in
Eq.~(\ref{eq:triple_integral_anisotropy}) for $P_{ai}$ and
Eq.~(\ref{eq:triple_integral}) for $\beta_{0(2)}$ and $\beta_{0(1)}$
and plotted the results in Fig.~\ref{fig:production_rate} with thick 
solid lines.

For $\sigma_{H}\ll 1$ and $\delta_{{\rm th}}\gg \sigma_{H}$, 
we have succeeded in deriving the following 
semianalytic expression for Eq.~(\ref{eq:triple_integral}):
\begin{equation}
\beta_{0}\simeq  
\frac{5\sqrt{5}\pi^{4}}{(2\cdot 3)^{9}}\bar{E}^{-5}\frac{\delta_{{\rm
th}}^{9}}{\sigma_{H}^{4}}\exp \left(-\frac{\delta_{{\rm
			       th}}^{2}}{2\sigma_{H}^{2}}\right),
\label{eq:semianalytic_anisotropy_threshold}
\end{equation}
where $\bar{E}\simeq 1.182$. The derivation of the above formula 
is described in Appendix~\ref{sec:derivation_formula}.
In Fig.~\ref{fig:production_rate}, we also plot with dashed lines 
the semianalytic formula~(\ref{eq:semianalytic_anisotropy}) for
$P_{{\rm ai}}$, Eq.~(\ref{eq:semianalytic_anisotropy_threshold}) with
$\delta_{{\rm th}}=\delta_{{\rm th}(2)}$ or
\begin{equation}
 \beta_{0(2)}\simeq  1.921\times 10^{-7} {\cal I}^{6}\sigma_{H}^{2}
\exp \left[-0.1474 \frac{{\cal I}^{4/3}}{\sigma_{H}^{2/3}}\right]
\label{eq:semianalytic_2nd}
\end{equation}
for $\beta_{0(2)}$, and Eq.~(\ref{eq:semianalytic_anisotropy_threshold}) with
$\delta_{{\rm th}}=\delta_{{\rm th}(1)}$ or
\begin{equation}
 \beta_{0(1)}\simeq 3.244\times 10^{-14}
  \frac{{q}^{18}}{\sigma_{H}^{4}}\exp\left[-0.004608
				    \frac{{q}^{4}}{\sigma_{H}^{2}}\right]
\label{eq:semianalytic_1st}
\end{equation}
for $\beta_{0(1)}$.
For $\beta_{0(2)}$,
Eqs.~(\ref{eq:semianalytic_2nd}) and (\ref{eq:semianalytic_anisotropy})
agree with the numerical result for $\sigma_{H}\lesssim 0.005$
and for $0.005\lesssim \sigma_{H}\lesssim 0.2$, respectively.
This means that angular momentum is more important for
$\sigma_{H}\lesssim 0.005$,
while anisotropic collapse is more important for 
$0.005\lesssim \sigma_{H}\lesssim 0.2$.
Also for $\beta_{0(1)}$, 
Eqs.~(\ref{eq:semianalytic_1st}) and (\ref{eq:semianalytic_anisotropy})
agree with the numerical result for $\sigma_{H}\lesssim 0.04$
and for $0.04\lesssim \sigma_{H}\lesssim 0.2$, respectively.

In Fig.~\ref{fig:production_rate}, 
we can also see that the suppression due to the 
second-order effect $\beta_{0(2)}$ is much weaker than that due to the 
first-order effect $\beta_{0(1)}$ for $\sigma_{H}\lesssim 0.02$. 
This means that if $\sigma_{H}\lesssim 0.02 $ and 
${q}$ is distributed around $0$, the
probability of black hole formation is 
dominated by masses with ${q}<{q}_{c}\simeq \sigma_{H}^{1/3}$.
Therefore, 
the assumption that the mass 
is initially given by a ball is naturally 
justified to estimate the effect of anisotropic collapse.
With the distribution of ${q}$ further taken into account, 
the probability of black hole formation is  
semianalytically estimated as
\begin{eqnarray}
 \beta_{0}\simeq \left\{
\begin{array}{cc}
1.921\times 10^{-7} f_{q}(q_{c}){\cal I}^{6}\sigma_{H}^{2}
\exp \left[-0.1474 \displaystyle\frac{{\cal I}^{4/3}}{\sigma_{H}^{2/3}}\right]
 & (\sigma_{H}\lesssim 0.005)\\
0.05556 \sigma_{H}^{5} & (0.005\lesssim \sigma_{H}\lesssim 0.2).
\end{array}
\right.,
\end{eqnarray}
where $f_{q}(q_{c})$ is the fraction of masses of which $q$ is smaller
than $q_{c}$.

For comparison, the production 
rate in the radiation-dominated phase, $P_{{\rm rd}}$, is also plotted 
with a thin solid line in this figure,
where the threshold is chosen to $\delta_{{\rm th}}=0.42$.
The production rate in the 
matter-dominated era is larger than that in the radiation-dominated 
phase for $\sigma_{H}\lesssim 0.05$, 
while they are comparable with each other
for $0.05\lesssim \sigma_{H}\lesssim 1$.

\subsection{Black hole threshold in the radiation-dominated era}
Here we review black hole threshold in the radiation-dominated phase 
in terms of density perturbation and curvature perturbation.
To define the curvature perturbation, we have to
introduce the 3+1 decomposition of the spacetime
\begin{equation}
 ds^{2}=-\alpha^{2}c^{2}dt^{2}+\psi^{4}a^{2}\tilde{\gamma}_{ij}(dx^{i}+\beta^{i}dt)(dx^{j}+\beta^{j}dt),
\end{equation}
where we choose $\tilde{\gamma}_{ij}$ so that its determinant equals to
that of the flat 3-metric. The curvature perturbation $\zeta$ is
defined as 
\begin{equation}
\zeta=-2\ln \psi
\label{eq:zeta}
\end{equation}
in the uniform-density slicing~\cite{Lyth:2004gb}. Some authors 
including Kopp {\it et al.}~\cite{Kopp:2010sh} take another sign on the 
right-hand side of Eq.~(\ref{eq:zeta}).
Primordial cosmological perturbations are given by 
long-wavelength solutions, where the length scale of 
the perturbation is much larger than the Hubble horizon 
scale~\cite{Lyth:2004gb}. 
In the lowest and second lowest orders of the long-wavelength limit, 
$\psi$ is time independent and of the order of unity. 
The density perturbation $\delta$ in the comoving slicing and the $\psi$ are
related to each other by the following relation~\cite{Harada:2015yda}:
\begin{equation}
 \delta=-\frac{4(1+w)}{3w+5}\frac{c^{2}}{a^{2}H^{2}}\frac{\Delta\psi}{\psi^{5}}
\label{eq:delta_psi}
\end{equation} 
for the equation of state $p=w\rho c^{2}$ with $w$ constant, 
where $\Delta$ is the
Laplacian of the flat 3-metric. This means 
\begin{equation}
 \tilde{\delta}:=\lim_{c/(aHr_{0})\to 0}\left(\frac{aHr_{0}}{c}\right)^{2}\delta
\end{equation}
is time independent, where $r_{0}$ is the comoving 
length scale of perturbation and is 
identified with $r_{0}$ in the previous sections. 
In general relativistic numerical simulations, $r_{0}$ has been 
chosen as the radius of 
the boundary of the overdense region.
The threshold for black hole formation has been 
discussed in terms of $\tilde{\delta}$ after it is averaged within $r_{0}$
and it can be identified with $\delta_{s,H}$ in the previous sections.
See~\cite{Peebles:1980,Hwang:2012bi} for the equivalence between the density
perturbations in Newtonian gravity and in the comoving slicing in 
general relativistic cosmological perturbation theory.

By recent numerical relativity 
simulations in spherical symmetry~\cite{Harada:2015yda}, 
the black hole threshold 
in the radiation-dominated era has been found and
is $\psi_{{\rm th}} \simeq 1.40-1.69$ in terms of the peak value of
$\psi$, which is equivalent to $|\zeta_{{\rm th}}|\simeq 0.67-1.05$ 
in terms of the peak value
through Eq.~(\ref{eq:zeta}), depending on the profile of the perturbation.
In terms of the density perturbation $\tilde{\delta}$,
the threshold is given by $\delta_{{\rm th}}\simeq 0.42-0.56$.
This result is fairly consistent with preceding works
~\cite{Shibata:1999zs,Musco:2004ak,Polnarev:2006aa,Musco:2008hv,Musco:2012au}.
The relation between $\tilde{\delta}$ and the peak value of $\psi$ 
(or $\zeta$) is not one to one but largely profile dependent.
In fact, 
$\delta_{{\rm th}}$ and $\psi_{{\rm th}}$ even show 
opposite behaviors on the sharpness of the transition between 
the overdense region and the flat Friedmann-Lema\^{i}tre-Robertson-Walker (FLRW) exterior, 
as can be seen in Tables I and II and Figs. 2 and 3 in~\cite{Harada:2015yda}.
This suggests that the black hole threshold is profile dependent 
because of the complexity of gravitational collapse against the pressure
gradient force.
The analytic formula for a gentle profile
is derived using a simple model of perturbation~\cite{Harada:2013epa}
and shows a good agreement with numerical results for $p=w\rho c^{2}$
with $0.01\le w \le
0.6$ obtained by Musco and Miller~\cite{Musco:2012au}. 
This formula gives $\tilde{\delta}\simeq 0.4135$ for the radiation 
fluid~\cite{Harada:2013epa}.

It should be noted that rather smaller values of $|\zeta_{{\rm th}}|$,  
$0.2131$ for the peak value and $0.0862$ for 
the averaged value, are 
 reported in Sec. IV of~\cite{Harada:2013epa}. 
These smaller values are 
due to the very special conversion function 
from $\tilde{\delta}$ to $\zeta$ given in~\cite{Kopp:2010sh}, which is based 
on the top-hat curvature profile.
In fact, as is shown in~\cite{Harada:2015yda}, 
this model contains an unphysical feature that
the density field has a negative
delta-functional term at 
the transition to the flat FLRW exterior and 
gives a considerably smaller value of $|\zeta|$ 
for the same $\tilde{\delta}$ than 
more physical models with smooth and non-negative 
density fields.

\subsection{Comparison between the matter-dominated and radiation-dominated eras}
It is useful to discuss the threshold in terms of the Fourier components of 
the curvature perturbation $\zeta$.
Linearizing Eqs.~(\ref{eq:zeta}) and 
(\ref{eq:delta_psi}), we obtain the relation 
\begin{equation}
\hat{\delta}_{{\bf k}}=-\frac{2(1+w)}{3w+5}\frac{c^{2}k^{2}}{(aH)^{2}}\hat{\zeta}_{{\bf k}},
\end{equation}
where $\hat{\delta}_{{\bf k}}$ and $\hat{\zeta}_{{\bf k}}$ are the 
Fourier components of $\delta(x^{i})$ and $\zeta(x^{i})$, respectively.
[A negative sign is missing on the right-hand side of
Eq.~(56) in~\cite{Harada:2016mhb}.]
We can express the averaged density perturbation in the real space
in terms of the Fourier components of $\zeta$ as  
\begin{equation}
\delta_{s,H}=\lim_{c/(aHr_{0})\to 0}\left(\frac{aHr_{0}}{c}\right)^{2}
\sum_{{\bf k}} \hat{\delta}_{{\bf k}}g(kr_{0})
=-\frac{2(1+w)}{3w+5}\sum_{{\bf k}}(kr_{0})^2 \hat{\zeta}_{{\bf
k}}g(kr_{0}).
\label{eq:delta_sH_zeta_hat}
\end{equation}
Thus, we find 
\begin{equation}
\sigma_{H}^{2}=\left[\frac{2(1+w)}{3w+5}\right]^{2}
\sum_{{\bf k}}(kr_{0})^4 \langle |\hat{\zeta}_{{\bf k}}|^{2}\rangle
g^{2}(kr_{0})
= 
\left[\frac{2(1+w)}{3w+5}\right]^{2}\int_{0}^{\infty}\frac{dk}{k}(kr_{0})^{4}
P_{\zeta}(k)W(kr_{0}),
\label{eq:sigma_H_P_zeta_W}
\end{equation}
where $W(kr_{0})$ is identified with $g^{2}(kr_{0})$.
We have assumed a random phase and isotropy in the distribution of
$\hat{\zeta}_{{\bf k}}$ and defined the power spectrum
$P_{\zeta}(k):=[k^{3}/(2\pi^{2})]\langle |\hat{\zeta}_{k}|^{2} \rangle$.
Because of the rapidly increasing function $k^{4}$ and the window function 
in Eq.~(\ref{eq:sigma_H_P_zeta_W}), the right-hand side can be written 
by the power spectrum at the characteristic wave number $k=k_{{\rm BH}}$.
Although there appears subtlety in identifying $k_{{\rm BH}}$,
we simply write Eq.~(\ref{eq:sigma_H_P_zeta_W}) as 
\begin{equation}
\sigma_{H}^{2}\simeq 
\left[\frac{2(1+w)}{3w+5}\right]^{2}
P_{\zeta}(k_{{\rm BH}}).
\label{eq:sigma_power_spectrum}
\end{equation}
Note that this agrees with
Eq.~(3.5) of Alabidi {\it et al.}~\cite{Alabidi:2013wtp} up to a factor of 2,
which will depend on the definition of $k_{{\rm BH}}$. This $\sigma_{H}$ can be directly compared with the threshold $\delta_{{\rm th}}$.  

Based on the above argument, let us compare the production rates
for the two eras.
In the radiation-dominated era, we 
find $\sigma_{H}^{2}\simeq (16/81)P_{\zeta}$.
From Eq.~(\ref{eq:delta_sH_zeta_hat}), we can have an approximate
relation
\begin{equation}
 \delta_{s,H}\simeq -\left.\frac{2(1+w)}{3w+5}\hat{\zeta}_{{\bf
		      k}}\right|_{k=k_{\rm BH}\simeq r_{0}^{-1}},
\label{eq:averaged_density_zeta_k}
\end{equation}
which is consistent with Eq.~(\ref{eq:sigma_power_spectrum}).
Therefore, the black hole threshold $\delta_{{\rm th}}\simeq 0.42-0.56$
in the radiation-dominated era 
is roughly equivalent to $ |\hat{\zeta}_{{\bf k}}|_{{\rm th}} \simeq 0.95-1.26$
by Eq.~(\ref{eq:delta_sH_zeta_hat}).
In the matter-dominated era, 
we have $\sigma_{H}^{2}=(4/25)P_{\zeta}$.
In the current paper, we find that 
for ${q}<{q}_{c}$, the threshold is given by  
$\delta_{{\rm th}}=(2{\cal I}\sigma_{H}/5)^{2/3}$, 
which is roughly equivalent to 
$|\hat{\zeta}_{{\bf k}}|_{{\rm th}} \simeq 0.74 {\cal
I}^{2/3}P_{\zeta}^{1/3}$
through Eqs.~(\ref{eq:sigma_power_spectrum}) and
(\ref{eq:averaged_density_zeta_k}), while for ${q}>{q}_{c}$, it
is given by $\delta_{{\rm th}}\simeq 0.096 {q}^{2}$, 
which is roughly equivalent to $|\hat{\zeta}_{{\bf k}}|_{{\rm th}} \simeq 0.24 {q}^{2}$.
If ${q}$ is distributed around $0$, 
we can conclude that black hole production is enhanced 
in the matter-dominated phase in comparison with 
the radiation-dominated phase because some 
fraction of masses have ${q}<{q}_{c}$ 
and those masses dominate the probability of black hole formation
and give a larger production rate than in the radiation-dominated era.
Even if ${q}<{q}_{c}$ is highly restricted, 
the masses with ${q}$ satisfying ${q}_{c}< {q}\lesssim
2.0-2.2$, 
which have the threshold value $|\hat{\zeta}_{{\bf k}}|_{{\rm
th}}$ smaller than that in the radiation-dominated era,
will dominate the production rate and give a larger production rate.

\section{Conclusion}
\label{sec:conclusion}

We conclude that angular momentum plays crucial roles in 
primordial black hole formation in the matter-dominated phase of 
the Universe if it lasts sufficiently long. 
In fact, the formation of primordial black holes 
is exponentially suppressed contrary to conventional expectations.
This suppression is much stronger than the effect of anisotropic
collapse and the conventional formula overestimates the production rate.
However, 
since the newly obtained exponential suppression is much weaker 
than that in the radiation-dominated era, the matter-dominated era can 
still be regarded as the epoch of enhanced production of primordial black holes.
We also find that most of the primordial 
black holes formed in the matter-dominated era were rapidly rotating 
at their formation epoch and still are if they have kept a large
fraction of spins until now.
This has interesting implications for astrophysics and cosmology.
We also predict that when primordial black holes are formed,
much more ``minihaloes'' are formed, which have supercritical 
values of the Kerr parameter.
If the matter-dominated era does not last so 
long, the production rate of primordial black holes is strongly
suppressed and the tendency towards large spins of both the  
primordial black holes and the minihaloes is 
significantly weakened. Since the duration of the
matter-dominated era is highly dependent on the cosmological
scenario, it would be very interesting from a cosmological point of view 
to investigate the finite duration effect on 
the spins of primordial black holes.

\acknowledgments

The authors are grateful to C.~Byrnes, B.~J.~Carr, J.~Garriga, 
T.~Hiramatsu, T.~Igata,
S.~Jhingan, T.~Kobayashi, I.~Musco, T.~Nakama, M.~Sasaki, T.~Suyama, T.~Tanaka, 
and S.~Yokoyama for helpful comments.
The authors also thank Tommi Tenkanen and Takahiro Terada.
This work was supported by JSPS
KAKENHI Grants No. JP26400282 (T.H.), No. JP16K17688, No. JP16H01097 (C.Y.), 
No. JP26247041, No. JP15H05889, No. JP16H0877, No. JP17H01131 (K.K.), and
No. JP25400265 (K.N.).

\appendix

\section{Exact expression for the first-order contribution in an ellipsoid}
\label{sec:exact_expression_ellipsoid}

First, we derive exact expressions for the integrals
given in Eq.~(\ref{eq:integral_formula}), where 
the region of integration is that 
inside the ellipsoid given by Eq.~(\ref{eq:ellipsoid}).
We replace the integration variable $\bx$ by $\by$ defined as
$
\bx=:(A_1y_1,A_2 y_2,A_3y_3)
$
and introduce a new wave number defined as
$
\tilde{\bk}:=(A_1k_1,A_2k_2,A_3k_3).
$
Then, we have
\begin{eqnarray}
&&\int_{V} e^{i\bk\cdot \bx}d^{3}\bx=A_1A_2A_3 \int_{|\by|\leq1}
 e^{-\tilde{\bk}\cdot\by}d^3\by=V g(\tilde{k}),\\
&&\int_V x_i e^{i\bk\cdot\bx} d^3\bx=A_1A_2A_3 A_i\int_{|\by|\leq1}y_i e^{-\tilde{\bk}\cdot\by}d^3\by
=\frac{i}{5}V\bar{k}_i f(\tilde{k}),
\end{eqnarray}
where 
$
\tilde{k}:=\sqrt{(A_1k_1)^2+(A_2k_2)^2+(A_3k_3)^2}
$
and 
$
\bar{\bk}:=(A_1^2k_1,A_2^2k_2,A_3^2k_3).
$
Hence, the averaged density perturbation and its variance are given by 
\begin{equation}
 \delta_{s}=\sum_{{\bf k}}A_{{\bf k}}g(\tilde{k})\quad \mbox{and}\quad 
 \langle \delta_{s}^{2}\rangle = \sum_{{\bf k}}\langle |A_{{\bf k}}|^{2}
  \rangle g^{2}(\tilde{k}),
\end{equation}
respectively, where $A_{{\bf k}}$ is assumed to take a random phase. 
We also have 
\begin{align}
\int_V \epsilon_{ijl}x_j \partial_j\psi_{l} d^3\bx=\epsilon_{ijl}\sum_{\bk}ik_l\hat{\psi}_\bk\int_V x_je^{i\bk\cdot\bx}d^3\bx
=\frac{2}{15}Va_0^2\epsilon_{ijl}\sum_{\bk}\frac{\bar{k}_j k_l}{k^2}f(\tilde{k})A_\bk.
\end{align}
The first-order contribution to the angular momentum ${\bf L}_{(1)}$ is then given by
\begin{equation}
L_{(1)i}=-\frac{2}{15}V\rho_0a^3ta_0^2\epsilon_{ijl}\sum_{\bk}\frac{\bar{k}_j k_l}{k^2}f(\tilde{k})A_\bk. \label{am}
\end{equation}
From Eq.~(\ref{am}), we have its variance
\begin{align}
\langle {\bf L}_{(1)}^2\rangle
&=\left(\frac{2}{3}\rho_0 a^3a_0^2t\right)^2\sum_{\bf k}\frac{f^2(\tilde{k})}{k^4}
\langle|A_{\bf k}|^2\rangle\cr
&\times \left[\left(i_1-i_2\right)^2\left(k_1k_2\right)^2
+\left(i_2-i_3\right)^2\left(k_2k_3\right)^2
+\left(i_3-i_1\right)^2\left(k_3k_1\right)^2\right]  \nonumber \\
&=(t \rho_{0}a^{3})^{2}\frac{4}{9}a_{0}^{4}\epsilon_{ijk}\epsilon_{ipq}
J_{km}J_{ql}\sum_{{\bf k}}f^{2}(\tilde{k})\frac{k_{j}k_{m}k_{p}k_{l}}{k^{4}}\langle
 |A_{{\bf k}}|^{2}\rangle .
\label{eq:exact_expression_L1}
\end{align}

If the eccentricity is low, or equivalently $q$ is small, 
we can neglect the anisotropy in 
$f^{2}(\tilde{k})$
and we have
\begin{equation}
\langle {\bf L}_{(1)}^2\rangle=\frac{2}{15}\left(\frac{2}{3}\rho_0 a^3a_0^2t\right)^2\left(\mu_1^2-3\mu_2\right)
\sum_{\bk}\langle|A_{\bf k}|^2\rangle f^2(kr_0),
\end{equation}
where the power spectrum is assumed to be isotropic.
However, if the eccentricity is high, 
or equivalently, $q$ is large, we cannot neglect the anisotropy in
$f^{2}(\tilde{k})$. Even in this case,  
however, we may still rewrite Eq.~(\ref{eq:exact_expression_L1}) 
in the following form:
\begin{equation}
\langle {\bf
 L}_{(1)}^2\rangle^{1/2}=
\frac{2}{5\sqrt{15}}
{\cal R}
{q}
\frac{MR^2}{t}\langle\delta^2_{\rm s}\rangle^{1/2},
\end{equation}
where 
\begin{align}
{\cal R}&:
=\sqrt{\frac{\epsilon_{ijk}\epsilon_{ipq}J_{km}J_{ql}\sum_{{\bf
 k}}f^{2}(\tilde{k})\displaystyle\frac{k_{j}k_{m}k_{p}k_{l}}{k^{4}}\langle |A_{\bf k}|^{2}\rangle}{\epsilon_{ijk}\epsilon_{ipq}J_{km}J_{ql}\sum_{{\bf
 k}}f^{2}(kr_{0})\displaystyle\frac{k_{j}k_{m}k_{p}k_{l}}{k^{4}}\langle |A_{\bf k}|^{2}\rangle}
\frac{\sum_{{\bf k}}f^{2}(kr_{0})\langle |A_{\bf k}|^{2}\rangle }
{\sum_{{\bf k}}g^{2}(\tilde{k})\langle |A_{\bf k}|^{2}\rangle }}.
\end{align}
${\cal R}$ does not depend on the overall normalization factor.
If ${\cal R}\simeq  1$, we recover Eq.~(\ref{eq:fluctuation_L1}).

\section{Condition for the end time in terms of the reheating temperature}
\label{sec:reheating_temperature}
As an interesting example, in this section we assume that 
the reheating process due to a decay of
massive particles makes the radiation-dominated phase start, i.e., 
$t_{{\rm end}}=t_{R}$, where $t_{R}$ is the cosmic time at the reheating.

In this case, for the wave number $k$ crossing the horizon ($k=aH$) at $t=t_H$
during the matter-dominated phase,  
the relation between $k$ and $t_{H}$ is given by 
\begin{equation}
  k \sim  k_{\rm eq}
     \left(
\frac{T_R}{T_{\rm eq}}
     \right)
     \left(
\frac{t_H}{t_{R}}
     \right)^{-1/3},
\end{equation}
where $k_{{\rm eq}}$ and $T_{{\rm eq}}$ are 
the wave number and the temperature at the (latest)
matter-radiation equality, respectively.
Then, we can show that Eq.~(\ref{eq:long_MD}) gives
an upper bound on the reheating temperature $T_R$ for the successful 
enhanced production of highly spinning primordial black holes as 
\begin{eqnarray}
 T_R \lesssim 0.1 {\rm GeV}
     \left(
\frac{k}{ 10^6 {\rm Mpc}^{-1} }
     \right)
     \left(
\frac{2}{5}{\cal I}\sigma_{H}
     \right)^{1/3},
\end{eqnarray}
where we have used the relation between 
$t_{R}$ and $T_{R}$ 
\begin{equation}
 t_{R}\simeq
  \left(\frac{g_{*}}{45/(2\pi^{2})}\right)^{-1/2}\frac{m_{{\rm
  Pl}}}{T_{R}^{2}}\sim \frac{m_{{\rm Pl}}}{T_{R}^{2}}
\end{equation}
with $g_{*}\simeq 10.75-106.75$ and 
put $m_{\rm Pl}\simeq 2.4 \times 10^{18} {\rm
GeV}$, $k_{\rm eq} \sim 0.01 {\rm Mpc}^{-1}$, and $T_{\rm eq} \sim 0.7
{\rm eV}$.

\section{Derivation of the semianalytic formula}
\label{sec:derivation_formula}

Changing the variables from $(\alpha,\beta,\gamma)$ to $(x,y,z)$ by 
\begin{equation}
 x=\frac{\alpha+\beta+\gamma}{3},\quad 
y=\frac{(\alpha-\beta)-(\beta-\gamma)}{4},\quad 
z=\frac{\alpha-\gamma}{2}
\end{equation}
and from $(x,y,z)$ to $(t,u,z)$ by 
\begin{equation}
 t=\frac{x}{z},\quad u=\frac{y}{z},
\end{equation}
we find that the distribution function for $(t,u,z)$ is given by  
\begin{equation}
\tilde{w}(t,u,z)dtdudz
=-\frac{27}{\sqrt{5}\pi\sigma_{3}^{6}}(2u-1)(2u+1)z^{5}\exp
\left[-A(t,u)z^{2}\right]dtdudz,
\end{equation}
where 
\begin{equation}
A(t,u):=\frac{9}{10}\left(\frac{t}{\sigma_{3}}\right)^{2}+2\left(\frac{u}{\sigma_{3}}\right)^{2}+\frac{3}{2}\left(\frac{1}{\sigma_{3}}\right)^{2}
\end{equation}
and the domain $\infty >\alpha \geq \beta \geq \gamma \geq -\infty$ is
transformed to $-\infty < x < \infty$, $-1/2<u<1/2$, and $0<z<\infty$.
Since
\begin{equation}
 h(\alpha,\beta,\gamma)=\tilde{h}(t,u,z)
:=\frac{4}{\pi z}\left(t+\frac{2}{3}u+1\right)^{-2}E\left(\sqrt{1-\left(u+\frac{1}{2}\right)^{2}}\right),
\end{equation}
the criterion for the black hole formation, $h<1$, is transformed to
\begin{equation}
 z>z_{*}(t,u):=\frac{4}{\pi}\left(t+\frac{2}{3}u+1\right)^{-2}\tilde{E}(u),
\end{equation}
where we put 
\begin{equation}
 \tilde{E}(u)=E\left(\sqrt{1-\left(u+\frac{1}{2}\right)^{2}}\right).
\end{equation}
If we take only the anisotropic collapse into account, we find 
\begin{equation}
 \beta_{0}=-\frac{27}{\sqrt{5}\pi \sigma_{3}^{6}}\int^{1/2}_{-1/2}du (2u-1)(2u+1)\int_{-1-(2/3)u}^{\infty}dt\int_{z_{*}(t,u)}^{\infty}dz~z^{5}\exp[-A(t,u)z^{2}].
\end{equation}
For $\sigma_{H}=\sqrt{5}\sigma_{3}\ll 1$, we can obtain 
the semianalytic formula
\begin{equation}
 \beta_{0}\simeq \frac{5\cdot 5^{3}\pi^{9/2}}{2^{9}\cdot
  3^{6}\sqrt{10}}\bar{E}^{-5}\sigma_{H}^{5}\simeq 0.05556\sigma^{5},
\end{equation}
where 
\begin{equation}
 \bar{E}^{-5}:=\frac{3}{2}\int_{-1/2}^{1/2}(1-2u)(1+2u)\tilde{E}^{-5}(u)
\label{eq:E_bar}
\end{equation}
and $\bar{E}\simeq 1.182$. The derivation of the above formula is given 
in Appendix B of \cite{Harada:2016mhb}.

To take both anisotropic collapse and angular
momentum into account, we would like to calculate the integral in  
Eq.~(\ref{eq:triple_integral}). This can be rewritten as
\begin{equation}
 \beta_{0}=-\frac{27}{\sqrt{5}\pi \sigma_{3}^{6}}\int^{1/2}_{-1/2}du
  (2u-1)(2u+1)\int_{-1-(2/3)u}^{\infty}dt\int_{z_{0}(t,u)}^{\infty}dz~z^{5}\exp[-A(t,u)z^{2}], 
\end{equation}
where $z_{0}(t,u):=\mbox{max}(z_{*}(t,u),z_{{\rm th}}(t))$ and $z_{{\rm
th}}(t):=\delta_{{\rm th}}/(3t)$.
We denote two roots of $z_{*}(t,u)=z_{{\rm th}}(t)$ with 
$t_{1}(u)$ and $t_{2}(u)$ ($t_{1}(u)<t_{2}(u)$). 
We can find 
\begin{equation}
 t_{1,2}(u)=\frac{18
				      \tilde{E}(u)-(3+2u)\pi\delta_{{\rm
				      th}}\mp 6
				      \sqrt{\tilde{E}(u)(9\tilde{E}(u)-(3+2u)\pi\delta_{{\rm
				       th}})}}{3\delta_{{\rm th}}},
\end{equation}
where $t_{1}(u)$ and $t_{2}(u)$ correspond to the upper and lower signs, respectively.
We have $z_{0}(t,u)=z_{*}(t,u)$ for $-1-(2/3)u<t<0$ and $t_{1}(u)<t<t_{2}(u)$, while 
$z_{0}(t,u)=z_{{\rm th}}(t)$ for $0<t<t_{1}(u)$ and $t_{2}(u)<t$.
The integration with respect to $z$ can be done explicitly using a
well-known formula, which is given by Eq.~(66)
in Appendix A of \cite{Harada:2016mhb}. The result is 
\begin{eqnarray}
 \beta_{0}&=&-\frac{27}{2\sqrt{5}\pi\sigma_{3}^{6}}\int_{-1/2}^{1/2}du
  (2u-1)(2u+1) \nonumber \\
  & &\times \left[\left(\int_{-1-(2/3)u}^{\infty}+\int_{t_{1}(u)}^{t_{2}(u)}\right)dt
	       F(A,z_{*})+\left(\int_{0}^{t_{1}(u)}+\int_{t_{2}(u)}^{\infty}\right)F(A,z_{{\rm
	       th}})\right],
\label{eq:t-integral}
\end{eqnarray}  
where 
\begin{equation}
 F(A,z):=\frac{2+2Az^{2}+A^{2}z^{4}}{A^{3}}\exp[-Az^{2}].
\end{equation}
The above expression is a result of exact transformation from 
Eq.~(\ref{eq:triple_integral}). 

Hereafter, we assume $\sigma_{3}\ll 1$ and $\delta_{{\rm th}}\gg
\sigma_{3}$ and look for an approximate
expression for $\beta_{0}$. 
From the behavior of $A(t,u)z_{0}^{2}(t,u)$, we can show that the dominant
contribution can come from the integral of the interval
$[t_{1}(u),t_{2}(u)]$ 
or $[t_{2}(u),\infty)$ on
the right-hand side of Eq.~(\ref{eq:t-integral}). 
We denote the former and latter contributions to $\beta_{0}$ with $I_{1}$
and $I_{2}$, respectively.
For $t\gtrsim 1$, we obtain
\begin{equation}
 A\simeq \frac{9t^{2}}{10\sigma_{3}^{2}},\quad z_{*}\simeq \frac{4}{\pi t^{2}}\tilde{E}(u)
\end{equation}
and, hence,
\begin{equation}
Az_{*}^{2}\simeq \frac{72\tilde{E}(u)}{\pi^{2}\sigma^{2}t^{2}},\quad 
 Az_{{\rm th}}^{2}\simeq \frac{\delta_{{\rm
  th}}^{2}}{10\sigma_{3}^{2}},\quad t_{2}(u)\simeq 
  \frac{12\tilde{E}(u)}{\pi\delta_{{\rm th}}},
\end{equation} 
where we have used $z_{*}(t_{2}(u),u)=z_{{\rm th}}(t_{2}(u))$.
Then, $I_{2}$ is calculated to give 
\begin{eqnarray}
I_{2} \simeq  
\frac{5\sqrt{5}\pi^{4}}{(2\cdot 3)^{9}}\bar{E}^{-5}\frac{\delta_{{\rm
th}}^{9}}{\sigma_{H}^{4}}\exp \left(-\frac{\delta_{{\rm
			       th}}^{2}}{2\sigma_{H}^{2}}\right),
\end{eqnarray}
where we have used $\delta_{{\rm th}}\gg \sigma_{H}$. 
On the other hand, we can estimate $I_{1}$ as
\begin{equation}
 I_{1}\sim \frac{\delta_{{\rm
  th}}^{7}}{\sigma_{H}^{2}}\exp\left(-\frac{\delta_{\rm th}^{2}}{2\sigma_{H}^{2}}\right)
\end{equation}
up to a numerical factor of the order of unity and this is clearly
subdominant to $I_{2}$.
Thus, we finally 
reach Eq.~(\ref{eq:semianalytic_anisotropy_threshold}).
The discussion does not change whether $\delta_{{\rm th}}=O(1)$ or
$\delta_{{\rm th}}=O(\sigma_{H}^{2/3})$.

\end{document}